\documentclass[fleqn,usenatbib]{mnras}

\usepackage{newtxtext,newtxmath}

\usepackage[T1]{fontenc}
\usepackage{bm}
\usepackage{cleveref}
\crefname{figure}{Fig.}{Figs.}
\crefname{table}{Table}{Tables}

\DeclareRobustCommand{\VAN}[3]{#2}
\let\VANthebibliography\thebibliography
\def\thebibliography{\DeclareRobustCommand{\VAN}[3]{##3}\VANthebibliography}

\usepackage{graphicx}	
\usepackage{amsmath}	
\usepackage{rotating}
\usepackage{xspace}
\usepackage{ulem}
\makeatletter 
  \patchcmd{\NAT@citex}
    {\@citea\NAT@hyper@{%
      \NAT@nmfmt{\NAT@nm}%
      \hyper@natlinkbreak{\NAT@aysep\NAT@spacechar}{\@citeb\@extra@b@citeb}%
      \NAT@date}}
    {\@citea\NAT@nmfmt{\NAT@nm}%
    \NAT@aysep\NAT@spacechar\NAT@hyper@{\NAT@date}}{}{}

  \patchcmd{\NAT@citex}
    {\@citea\NAT@hyper@{%
      \NAT@nmfmt{\NAT@nm}%
      \hyper@natlinkbreak{\NAT@spacechar\NAT@@open\if*#1*\else#1\NAT@spacechar\fi}%
        {\@citeb\@extra@b@citeb}%
      \NAT@date}}
    {\@citea\NAT@nmfmt{\NAT@nm}%
    \NAT@spacechar\NAT@@open\if*#1*\else#1\NAT@spacechar\fi\NAT@hyper@{\NAT@date}}
    {}{}
\makeatother

\newcommand{\msun}{{\,\rm M_\odot}}

\newcommand{\cMpc}{{\rm cMpc}}
\newcommand{\kpc}{{\rm kpc}}

\newcommand{\HI}{\ion{H}{I}\xspace}
\newcommand{\HeI}{\ion{He}{I}\xspace}
\newcommand{\HeII}{\ion{He}{II}\xspace}

\newcommand{\hii}{\ion{H}{II}\xspace}
\newcommand{\hi}{\ion{H}{I}\xspace}
\newcommand{\thesan}{\textsc{thesan}\xspace}

\newcommand{\thzoom}{\mbox{\textsc{thesan-zoom}}\xspace}
\newcommand{\thesandarkone}{\mbox{\textsc{thesan-dark-1}}\xspace}
\newcommand{\thesanone}{\mbox{\textsc{thesan-1}}\xspace}

\newcommand{\HM}{\ion{H}{$_2$}\xspace}
\newcommand{\HII}{\ion{H}{II}\xspace}
\newcommand{\HeIII}{\ion{He}{III}\xspace}

\title[Population III stars in \thzoom]{The \thzoom project: Population III star formation continues until the end of reionization}

\author[Zier et al.]{%
Oliver Zier,$^{1, 2}$\thanks{E-mail: \href{mailto:oliver.zier@cfa.harvard.edu}{oliver.zier@cfa.harvard.edu}}
Rahul Kannan,$^{3}$
Aaron Smith,$^{4}$
Ewald Puchwein,$^{5}$
Mark Vogelsberger,$^2$
Josh Borrow,$^{6}$
\newauthor
Enrico Garaldi,$^{7,8}$
Laura Keating,$^{9}$
William McClymont,$^{10, 11}$
Xuejian Shen,$^2$
and Lars Hernquist$^1$
\\
$^{1}$ Center for Astrophysics | Harvard \& Smithsonian, 60 Garden St, Cambridge, MA 02138, USA\\
$^{2}$ Department of Physics, Kavli Institute for Astrophysics and Space Research, Massachusetts Institute of Technology, Cambridge, MA 02139, USA\\
$^{3}$ Department of Physics and Astronomy, York University, 4700 Keele Street, Toronto, ON M3J 1P3, Canada \\
$^4$ Department of Physics, The University of Texas at Dallas, Richardson, TX 75080, USA \\
$^5$ Leibniz-Institut f\"ur Astrophysik Potsdam, An der Sternwarte 16, 14482 Potsdam, Germany \\
$^6$ Department of Physics and Astronomy, University of Pennsylvania, 209 South 33rd Street, Philadelphia, PA 19104, USA \\
$^7$ Kavli IPMU (WPI), UTIAS, The University of Tokyo, Kashiwa, Chiba 277-8583, Japan \\
$^8$ Institute for Fundamental Physics of the Universe, via Beirut 2, 34151 Trieste, Italy \\
$^9$ Institute for Astronomy, University of Edinburgh, Blackford Hill, Edinburgh, EH9 3HJ, UK \\
$^{10}$ Kavli Institute for Cosmology, University of Cambridge, Madingley Road, Cambridge CB3 0HA, UK\\
$^{11}$ Cavendish Laboratory, University of Cambridge, 19 JJ Thomson Avenue, Cambridge CB3 0HE, UK\\
}

\date{Accepted XXX. Received YYY; in original form ZZZ}

\pubyear{2025}

\begin{document}
\label{firstpage}
\pagerange{\pageref{firstpage}--\pageref{lastpage}}
\maketitle

\begin{abstract}
Population III (Pop III) stars are the first stars in the Universe, forming from pristine, metal-free gas and marking the end of the cosmic dark ages. Their formation rate is expected to sharply decline after redshift $z \approx 15$ due to metal enrichment from previous generations of stars. 
In this paper, we analyze 14 zoom-in simulations from the \thzoom project, which evolves different haloes from the \thesanone cosmological box down to redshift $z=3$.
The high mass resolution of up to $142\msun$ per cell in the gas phase combined with a multiphase model of the interstellar medium (ISM), radiative transfer including Lyman-Werner radiation, dust physics, and a non-equilibrium chemistry network that tracks molecular hydrogen, allows for a realistic but still approximate description of Pop III star formation in pristine gas.
Our results show that Pop III stars continue to form in low-mass haloes ranging from $10^6 \msun$ to $10^9\msun$ until the end of reionization at around $z=5$. At this stage, photoevaporation suppresses further star formation in these minihaloes, which subsequently merge into larger central haloes. Hence, the remnants of Pop III stars primarily reside in the satellite galaxies of larger haloes at lower redshifts. 
While direct detection of Pop III stars remains elusive, these results hint that lingering primordial star formation could leave observable imprints or indirectly affect the properties of high-redshift galaxies. Explicit Pop III feedback and specialized initial mass function modelling within the \thzoom framework would further help interpreting emerging constraints from the \textit{James Webb Space Telescope}.
\end{abstract}




\section{Introduction}
The decoupling of the Cosmic Microwave Background (CMB) radiation field from the baryonic matter at the time of recombination, around redshift $z \approx 1100$, allowed small overdensities to grow and collapse in the subsequent cosmic dark ages.
The baryonic structure formation process was accelerated by the deeper gravitational potential wells created by pre-existing dark matter overdensities, which grew in a hierarchical, bottom-up fashion. 
The first collapsed objects were so-called ``minihaloes'' with masses below $\lesssim 10^{7}\,\msun $ \citep{ hartwig2022public}, which are thought to be the birthplaces of the first stars emerging around $z\approx 30$ \citep{klessen2023first}. 
The first generation of stars, known as Population III (Pop III) stars, is characterised by their formation from pristine gas that only contains the elements formed shortly after the Big Bang, or more precisely, the absence of metals except for lithium, i.e. metal-free ``primordial`` stars.
Their birth represents the end of the cosmic dark ages and the onset of the production of new electromagnetic radiation.

As galaxy assembly progresses, ``atomic cooling haloes`` with virial temperatures above $8,000$\,K form, which can efficiently cool gas by atomic hydrogen \citep{sutherland1993cooling}.
Pristine gas can reach lower temperatures through molecular hydrogen, H$_2$, which can cool through the excitation of its vibrational states.
However, this cooling mechanism becomes inefficient below $200$\,K \citep{greif2015numerical}, and reaching lower temperatures is only possible under special conditions in very massive or externally irradiated haloes that allow for efficient formation of deuterated hydrogen (HD) \citep{nagakura2005formation, Glover2008}.
These temperatures are significantly higher than the typical $10$\,K achieved by metal line cooling in molecular clouds in the present-day Universe \citep{gnedin2016physical}.
As a result, the Jeans mass required for the formation of Pop III stars is significantly higher, which reduces fragmentation and leads to a more top-heavy initial mass function (IMF) compared to the IMF observed at the present day \citep{salpeter1955luminosity,kroupa2002initial, chabrier2003galactic}.
It is important to note that forming less massive stars is still possible, e.g., through disk fragmentation in early accretion disks \citep[e.g.][]{greif2011simulations,clark2011formation,greif2012formation, stacy2013constraining,susa2019merge,jaura2022trapping, chiaki2022disc, prole2022fragmentation} or due to turbulence at the cloud level \citep{turk2009formation,stacy2010first,clark2011formation}, but the overall consensus is that Pop III stars generally have higher masses.

As with all stars, the subsequent evolution and ultimate fate of a Pop III star depends strongly on its initial mass. 
A massive star can end its life within a few million years as a core-collapse supernova or as a more energetic pair-instability supernova \citep[PISN,][]{fowler1964neutrino,heger2002nucleosynthetic,woosley2017pulsational,farmer2019mind}. 
Supernovae inject metals into their host halo and in the case of smaller haloes or highly energetic supernovae, they can even disperse metals into the surrounding intergalactic medium (IGM) and nearby minihaloes. This metal enrichment allows for the formation of the next generation of stars, which can cool more efficiently using the newly available metals \citep{kitayama2005supernova,greif2010first, ritter2012confined, jeon2014recovery,jaacks2018baseline,chiaki2019seeding,magg2022metal}.
However, there are certain mass ranges \citep[e.g. between $40\msun$ and an uncertain upper limit of  $70\msun - 100\msun$, ][]{klessen2023first} in which it is believed no supernova occurs and the star collapses directly to a black hole \citep{heger2002nucleosynthetic}.
The absence of metals prevents Pop III stars from generating strong stellar winds \citep{kudritzki2002line, krtivcka2006winds, krtivcka2009cno}. However, they produce large amounts of hydrogen ionizing radiation \citep{hartwig2022public} and less energetic Lyman-Werner (LW) radiation \citep[11.2eV - 13.6 eV,][]{kitayama2004structure,schauer2015lyman,schauer2017lyman}.
This LW radiation is particularly important because it can penetrate the IGM \citep{haiman2000radiative} and create a radiation background while dissociating molecular hydrogen. This dissociation can prevent gas cooling in distant minihaloes \citep{haiman1997destruction,ahn2009inhomogeneous}. 
In addition to reducing the star formation rate of Pop III stars, this process could potentially lead to the formation of supermassive stars with masses of up to $10^5\msun$ seeding the first massive black holes \citep{wise2008resolving, Smith2019, inayoshi2020assembly, Reinoso2023, Kiyuna2024}.

Pop III stars are extremely difficult to detect because their formation density peaks around redshift $z \approx 15 -20$, as even a small amount of metals and dust allow more efficient cooling and thus ``normal'' Population II (Pop II) star formation begins to dominate \citep{johnson2013first,de2014probing, sarmento2019following, liu2020did}.
Even a massive $1000\, \mathrm{M}_\odot$ star at $z> 10$ would be at least two orders of magnitude too faint to be observed by the \textit{James Webb Space Telescope} \citep[\textit{JWST};][]{schauer2020ultimately}.
However, strong gravitational lensing could amplify the signal \citep{rydberg2013detection,diego2019universe}.
For instance, the lensed star Earendel at $z=6.2$ is an unlikely yet potential candidate for a Pop III star \citep{welch2022jwst, schauer2022probability}.
Alternatively, Pop III PISNe occurring within the range $7 < z < 15$ should be bright enough to be observed by JWST \citep{hummel2012source,magg2016new,hartwig2018detection,venditti2024first}, but their short lifetimes result in a limited number of expected observations and uniquely identifying them remains an open question.
Indirect detection methods, such as examining the imprint of Pop III stars on the 21\,cm signal of neutral hydrogen  \citep{madau2014cosmic,bowman2018absorption,mesinger2019cosmic,magg2022effect,gessey2022impact, cruz2024first}, analyzing the metal abundances in extremely metal-poor stars believed to be their direct descendants \citep{frebel2005nucleosynthetic,salvadori2019probing, placco2021splus,skuladottir2021zero, Aguado2023} or gravitational waves caused by mergers of the remnants of Pop III stars \citep{Dayal2019, Neijssel2019, Tang2020, Ng2021, Tanikawa2021, Tanikawa2022}, are promising approaches for studying Pop III stars.

Modelling Pop III star formation and feedback is notoriously challenging because minihaloes below $10^7\msun$ require extremely high resolution \citep{klessen2023first}.
Therefore, most of our knowledge about Pop III stars comes from small-volume ``zoom-in'' simulations that aim to resolve their complete formation and growth processes. 
Furthermore, in addition to modelling cosmic expansion, gravitational forces, hydrodynamics, and cooling, other physical ingredients are needed.
For example, magnetic fields may play a role by reducing the fragmentation \citep{peters2014low, sharda2021magnetic,prole2022primordial,stacy2022magnetic,saad2022impact,sharda2024population, Sadanari2024}, though they can be weakened by nonideal effects \citep{mckee2020magnetic, sadanari2023non,Mayer2025}. 
Detailed chemical networks that also include deuterium and  H$^-$\citep{glover2005formation, klessen2023first, Nishijima2024}, are necessary. One may also need to take into account the streaming velocity between dark matter and baryonic matter in the early Universe \citep{tseliakhovich2010relative,greif2011delay,schauer2019influence, schauer2021influence,lake2024supersonic} and radiation transport that includes the Lyman-Werner band \citep{haiman1997destruction,skinner2020cradles,kulkarni2021critical,schauer2021influence,patrick2023collapse, Sugimura2024}.
To determine the final IMF, these simulations must be run until the accretion stops due to internal feedback processes \citep{Hosokawa2011, Sugimura2020, chon2024impact}.
These studies find that for gas-dominated cooling, a critical metallicity of  $10^{-3}$ to $10^{-4}\,\text{Z}_\odot$ is necessary to achieve a modern universal IMF \citep{omukai2000protostellar, bromm2001fragmentation, maio2010transition, maio2011interplay}. In contrast, for the dust-driven transition favoured by current observations, a metallicity of $10^{-6} - 10^{-4}\,\text{Z}_\odot$  is sufficient for producing Pop II stars \citep{schneider2002first, schneider2006constraints, schneider2012first, schneider2012formation, omukai2005thermal, chiaki2014dust}.
Nevertheless, there are indications that the IMF only shifts toward the present IMF when metallicity exceeds  $Z > 10^{-2}\,\text{Z}_\odot$ and may also depend on the redshift due to heating from the CMB \citep{chon2021transition, chon2022impact,chon2024impact}.

Large-scale cosmological simulations can complement small-scale simulations, which are used to study the IMF and formation of individual Pop III stars.
These larger simulations can model the full Epoch of Reionization (EoR) of the IGM ending at redshifts of $z\approx 5-6$ \citep{fan2006constraining, becker2015evidence,bosman2022hydrogen}. They also allow for a statistical analysis of the density of Pop III star formation as a function of redshift.
However, these larger simulations typically have significantly worse resolution, preventing them from resolving the smallest minihaloes that are capable of forming Pop III stars.
Additionally, only some simulations include an explicit Pop III subgrid model, which is typically based on the metallicity of newly formed star particles, while others apply the same feedback prescriptions used for Pop II stars. These subgrid models should be viewed as best efforts and still require calibration due to insufficient resolution.
Furthermore, many studies do not use radiative transfer for the Lyman-Werner band or include molecular hydrogen chemistry, reducing their predictability.
Examples of studies that analyse the formation of low metallicity stars include {\small THESAN-HR} \citep{borrow2023thesan,shen2024thesan}, simulations with dustyGadget \citep{di2023assembly, venditti2023needle, venditti2024first}, FLARE \citep{lovell2021first, vijayan2021first, wilkins2023first}, the Renaissance
simulations \citep{o2015probing, xu2016late}, the GIZMO simulations from \cite{jaacks2019legacy} and \cite{liu2020did}, and the RAMSES simulations from \cite{pallottini2014simulating}, \cite{sarmento2018following} and \cite{sarmento2022effects}.
Almost all of these simulations predict the formation of Pop III stars up to the end of the EoR. 
Interestingly, the TNG50 simulation \citep{pillepich2018first,nelson2019first} predicts the formation of Pop III stars even down to a redshift of $z=0$ \citep{pakmor2022formation}. However, as the resolution increases, the number of star-forming haloes that consist of purely pristine gas at $z=0$ decreases, suggesting that this effect may completely disappear at sufficiently high resolutions and is an artefact of the effective equation of state used in the study. 

This paper is part of the \thzoom project \citep[introduced in ][]{zoomIntro}, which consists of zoom-in simulations of fourteen haloes spanning the halo mass range from $10^8 \, \mathrm{M}_\odot$ to $10^{13} \, \mathrm{M}_\odot$ at the final redshift $z=3$. It includes radiative transfer with seven frequency bands, including Lyman-Werner radiation as well as the large-scale background radiation field from the parent \thesanone box \citep{Thesan1, ThesanAaron, ThesanEnrico, ThesanDR} as boundary condition at the edge of the high-resolution region.
Additionally, it includes dust and molecular hydrogen chemistry, and its dark matter resolution with particle masses down to $762 \, \mathrm{M}_\odot$ per resolution element allows us to resolve minihaloes of $10^6 \, \mathrm{M}_\odot$ by at least 1,000 particles for the smallest target halos.
This corresponds to the smallest haloes capable of cooling and forming Pop III stars \citep{glover2012first}.
For the largest target halos with a lower resolution of $4.86 \times 10^4\msun$ per dark matter particle, we still resolve larger minihalos of $10^7 \, \mathrm{M}_\odot$ with 200 dark matter particles.
While we do not implement specialized Pop III feedback or an explicit top-heavy IMF, our criterion that purely pristine gas forms Pop III stars captures first-order aspects of primordial star formation in a cosmological context, making \thzoom particularly well-suited for analyzing the formation of Pop III stars during the epoch of reionization.
 \thzoom was also used to study the star formation efficiency \citep{zoomJacob, Wang2025}, the star-forming main sequence \citep{McClymont2025a}, galaxy sizes \citep{McClymont2025b}, and the impact of external reionisation on galaxy properties \citep{zier2025}.

This paper is structured as follows:
In \cref{sec:simulations}, we summarize the \thzoom project and our analysis methods. 
In \cref{sec:stellarMetallicityDistribution}, we present the stellar metallicity distribution as a function of age at redshift $z = 3$.
In \cref{sec:PopIIIStarFormation}, we show that star-forming, pristine gas can mainly be found in minihaloes and satellite galaxies and that most young Pop III stars form ex-situ in subhaloes of mass $10^7 \, \mathrm{M}_\odot$ to $10^8 \, \mathrm{M}_\odot$.
We discuss and compare the results with the existing simulations and observations in \cref{sec:discussion} and summarise them in \cref{sec:summary}.

\section{Methods}
\label{sec:simulations}
\subsection{The \thzoom project}
\label{subsec:methods}
\begin{figure*}
    \centering
    \includegraphics[width=1\linewidth]{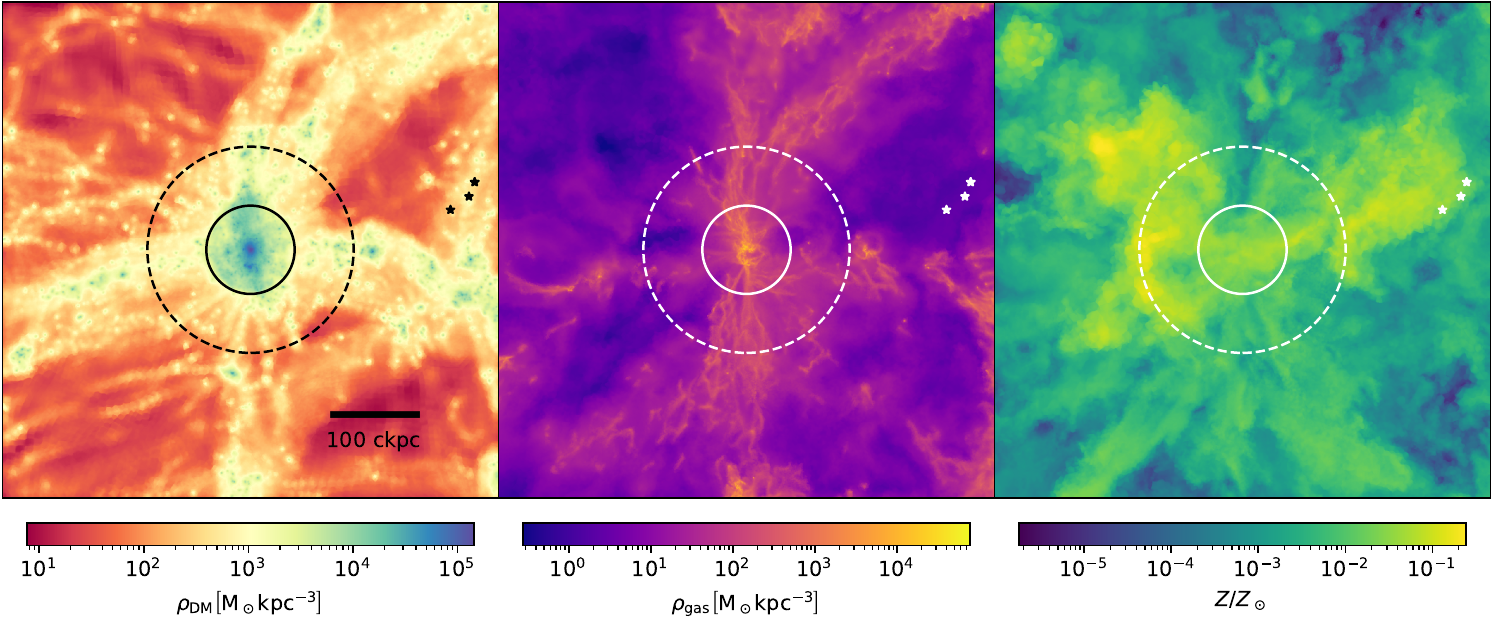}
    \caption{The dark matter density (left), gas density (middle), and gas metallicity (right) centred around the target halo m10.8\_8x at redshift $z=7$. We show volume averaged quantities in a slice in z-direction of depth $\rm 50\,ckpc$ and the comoving virial radius at $z=7$ (circle with solid line) and for comparison the comoving virial radius at $z=3$ (circle with dashed line). The star symbols represent Pop III star particles that formed between $z=7$ and $z=8$.
    All Pop III stars formed outside of the central halo but can merge later with it. The gas around the central halo almost reaches solar metallicity, but there are still pockets of metal-free gas outside of the halo.}
    \label{fig:51_cosmo_new_bin}
\end{figure*}

 \begin{table}
	\centering
 \small\addtolength{\tabcolsep}{-1.0pt}
	\begin{tabular}{llcccccccc} 
		\hline
		Halo  & $M_{\rm halo}$& $m_\mathrm{DM}$ & $m_\mathrm{gas}$ & $\epsilon_\mathrm{DM, stars}$ & $\epsilon_\mathrm{gas}^\mathrm{min}$\\  
		&  [$\mathrm{M}_\odot$] &[$\mathrm{M}_\odot$]& [$\mathrm{M}_\odot$] & [cpc] & [cpc]\\
		\hline
		m9.3\_16x  &$2.0 \times 10^{9}$& $7.62 \times 10^2$ & $ 1.4 \times 10^2$ & $138.30$ & $17.3$\\
		m10.8\_8x  &$5.9 \times 10^{10}$& $6.09 \times 10^3$ & $1.1 \times 10^3$ & $276.79$ & $34.6$ &\\
		m12.6\_4x   &$4.1 \times 10^{12}$& $4.86 \times 10^4$ & $9.1 \times 10^3$ & $553.59$ & $69.2$\\
		\hline
	\end{tabular}
    \caption{The main simulations we analyze in this paper: From left to right, the columns indicate the name of the simulation, the mass of the target halo at $z=3$ in the dark matter only simulation \thesandarkone, the mass of the high-resolution dark matter and gas particles and the (minimum) softening length of (gas) star and dark matter particles.
    We show in \cref{app:additionalPlots} results for eleven additional simulations that qualitatively agree with the three simulations discussed in the main part of this paper.}
    	\label{table:res}
\end{table}
The \thzoom suite is introduced in \cite{zoomIntro}, which provides a detailed discussion of the numerical methods and an overview of all simulations. This section briefly summarizes the most important components, particularly those essential for the formation of Pop III stars. 
The 14 dark matter (DM) haloes are drawn from the DM-only simulation \thesandarkone, which uses the same initial conditions as the flagship \thesanone simulation from the \thesan project \citep{Thesan1, ThesanAaron, ThesanEnrico, ThesanDR}.
The latter combined the successful IllustrisTNG galaxy formation model \citep{Weinberger2017, Pillepich2018Model}, which is derived from the Illustris model \citep{Vogelsberger2013, IllustrisNature, IllustrisIntro, genelIllustris}, with radiative transfer on the fly.
The target haloes were selected at redshift $z=3$ and cover a nearly uniform range in log space for halo masses from $10^8\msun$ to $10^{13}\msun$.
The high-resolution regions cover a sphere with a radius of about four times the virial radius of the central halo at $z=3$.
 This condition implies that, at high redshift, the high-resolution regions cover a considerably larger volume than just the immediate surroundings of the central haloes.

All simulations performed within the \thzoom project employ the massively parallel {\small AREPO} code \citep{springel2010pur,pakmor2016improving,weinberger2020arepo}, which solves the Euler equations on a moving, unstructured Voronoi mesh using a second-order accurate finite volume scheme.
The quasi-Lagrangian nature of the method results in an approximately constant mass resolution in the gas phase, which is further enforced by removing (splitting) cells that deviate from a preset target mass $m_{\rm gas}$ by more than a factor of 0.5 (2).
Gravitational forces are calculated using the hybrid TreePM method \citep{Bagla2022}, which uses a hierarchical octree \citep{Barnes1986} to compute short-range forces and the efficient particle mesh method for long-range forces \citep{Aarseth2003}.
The code also solves hyperbolic conservation laws for both the zeroth and first moments of the radiative intensity \citep{Kannan2019}, i.e. the photon number density and photon flux, along with the M1 closure relation \citep{Levermore1984, Dubroca1999}.
To enhance the scalability of the radiative transfer solver, a new communication pattern introduced in \cite{zier2024adapting} is used.
The radiation field is discretized into seven frequency bins: infrared (IR, $0.1-1$ eV), optical ($1.0-5.8$ eV), far-UV ($5.8-11.2$ eV), Lyman-Werner (LW, $11.2-13.6$ eV), the hydrogen ionizing band ($13.6-24.6$ eV), and two helium ionizing bands ($24.6-54.4$ eV, $>54.4$ eV). 

The radiative transfer solver is coupled to a non-equilibrium chemical network \citep{Kannan2020b} which evolves the abundances of the primordial species $\HM, \HI, \HII, \HeI, \HeII,$ and $\HeIII$.
We provide a more detailed discussion of the molecular hydrogen formation reactions in \cref{app:molecularHydrogen}, given their critical role in Pop~III star formation, and elaborate on the assumptions underlying our chemical network.
Non-equilibrium cooling and heating rates from these species are calculated self-consistently and used in an implicit time integration scheme \citep{Kannan2019}.
Metal line cooling is incorporated using cooling tables that assume ionization equilibrium with the UV background from \cite{FG09}
and is scaled linearly with the gas metallicity \citep{Vogelsberger2013}.
Additionally, the code accounts for photoelectric heating from Far-UV photons \citep[$5.8-11.2 \,  \mathrm{eV}$, based on][]{Wolfire2003}, cooling from gas-dust interactions \citep[based on][]{Burke1983} and Compton cooling/heating due to the CMB.
Cosmic dust is treated as an additional scalar property of gas cells. This representation neglects the relative velocities between gas and dust \citep{McKinnon2016, McKinnon2017}.
The dust is produced by supernovae and winds from asymptotic giant branch (AGB) stars, grows in the dense ISM and is destroyed by SN shocks and sputtering.
It interacts with the infrared radiation bin (IR, $0.1-1$ eV) and modifies the cooling and heating rates in the gas phase \citep{Kannan2021}.

For star formation and stellar feedback through stellar radiation, stellar winds and supernovae explosions, we use a significantly evolved version of the SMUGGLE model \citep{Marinacci2019}.
To be eligible for star formation, a gas cell must be smaller than its thermal Jeans length and have a density greater than $n_\mathrm{H} = 10~\mathrm{cm}^{-3}$.
From cells meeting both criteria we generate collisionless stellar particles using a standard probabilistic method with a star formation efficiency of 100\%  per free-fall time.
This high value prevents unresolved gas from collapsing to artificially high densities \citep{Hu2023}, but we note that the halo-scale star formation efficiency is quite robust to changes in resolution and the cell-based star formation efficiency \citep[for a detailed discussion see][]{zoomJacob}.
In the ISM the criterion of being unstable according to the Jeans criterium is the more strict one, as we show in \cref{app:densityDistribution}.
Cells with mass below twice the target gas mass $m_{\rm gas}$ are fully converted into one stellar particle; otherwise, only $2 m_{\rm gas}$ is removed from the gas cell to form the new particle.
We assume a Chabrier Initial Mass Function \citep[IMF, ][]{chabrier2003galactic} with minimum and maximum masses of $0.1~\mathrm{M}_\odot$ and $100~\mathrm{M}_\odot$.
Age and metal-dependent luminosities are calculated using the Binary Population and Spectral Synthesis models (BPASS; \citealt{BPASS2017}).
The radiation is injected in the 16 nearest gas cells across seven frequency bins, without tracking the spectral shape within individual bins. Instead, averaged properties such as ionization cross-sections per bin are calculated using a $2$\,Myr spectrum at quarter solar metallicity.
To reduce the impact of underresolved  \hii regions, we apply the corrections described in \cite{deng2024}.
For the mass loss rate and energy of stellar winds from young OB stars and AGB stars, we employ analytic prescriptions from \cite{Hopkins2018}, which are based on the \textsc{Starburst99}~\citep{Leitherer1999} stellar evolution model.
Metal enrichment rates follow \cite{Vogelsberger2013} though we note that there is a floor of $Z = 2.7\times10^{-9}$ in the initial conditions.
Stars with masses greater than $8\msun$ explode as supernovae, injecting $10^{51}~\mathrm{ergs}$ of energy into the ISM. To avoid the overcooling problem \citep{1Katz1996, DV2012} caused by not resolving the Sedov-Taylor phase, we instead inject the terminal momentum at the transition to the momentum-conserving phase \citep{Marinacci2019}.
The time of each supernova event is determined stochastically based on the IMF.
To better match the stellar-mass-halo-mass relations at high redshift \citep[e.g.][]{Moster2018, Behroozi2019}, an additional feedback channel is used to inject momentum during the first $5~\mathrm{Myrs}$ until the first SN events occur. This `Early Stellar Feedback' may represent missing physical processes and is further motivated in \cite{zoomIntro}.

A novel feature of the \thzoom project is its self-consistent modelling of the influence of external radiation sources outside the zoom-in region. 
This external radiation can dominate the radiation field around small haloes, leading to reduced inflows and lower star formation rates \citep[e.g.][]{Rees1986, Shapiro2004, Okamoto2008}.
While reionization is known to be patchy rather than uniform, most cosmological simulations typically employ a redshift-dependent, spatially uniform UV background \citep[UVB, e.g.][]{FG09, Haardt2012}, combined with a purely density-based self-shielding description \citep[e.g.][]{Rahmati2013}.
As shown in \cite{borrow2023thesan}, this neglect of the patchiness of reionization significantly affects the properties of low-mass galaxies and prevents the formation of metal-free stars at redshifts below $z=9$.
In the standard simulations of the \thzoom project, we use the evolving UV radiation field from the parent \thesanone simulation as a time-dependent boundary condition in the low-resolution region.
Below $z=5.5$, which marks the end of the \thesanone simulation, a smooth transition to the UVB from \cite{FG09} in the low-resolution region is implemented. 
Even in this case, self-shielding in the high-resolution region is done self-consistently.
The \thesanone simulation only evolved the hydrogen ionising and two helium ionising bands; therefore, our simulations do not incorporate a cosmological Lyman-Werner background, which is expected to form \citep{haiman2000radiative, incatasciato2023modelling}, but only local sources.
In a companion paper \citep{zier2025}, we analyse in detail the effects of this improved treatment of external radiation compared to the standard uniform UVB prescription for the \thzoom galaxies. 

In this paper, we focus on the default physics runs of the 14 target haloes, which are simulated using different resolution factors, as presented in \cref{table:res}. 
We use the default naming scheme introduced in the introduction paper \citep{zoomIntro}, which uses the halo mass at redshift $z = 3$ from the original \thesandarkone simulation along with the spatial zoom factor relative to the parent simulation. 
In the main body of this paper, we will concentrate on the simulations m12.6\_4x, m10.8\_8x, and m9.3\_16x as representative models for different halo masses. Results from the other simulations will be shown in \cref{app:additionalPlots}.
We note that all results presented in the main part of the paper agree qualitatively with those presented in the Appendix.

\subsection{Analysis methods and Pop III selection criteria}
\begin{figure*}
    \centering
    \includegraphics[width=0.65\linewidth]{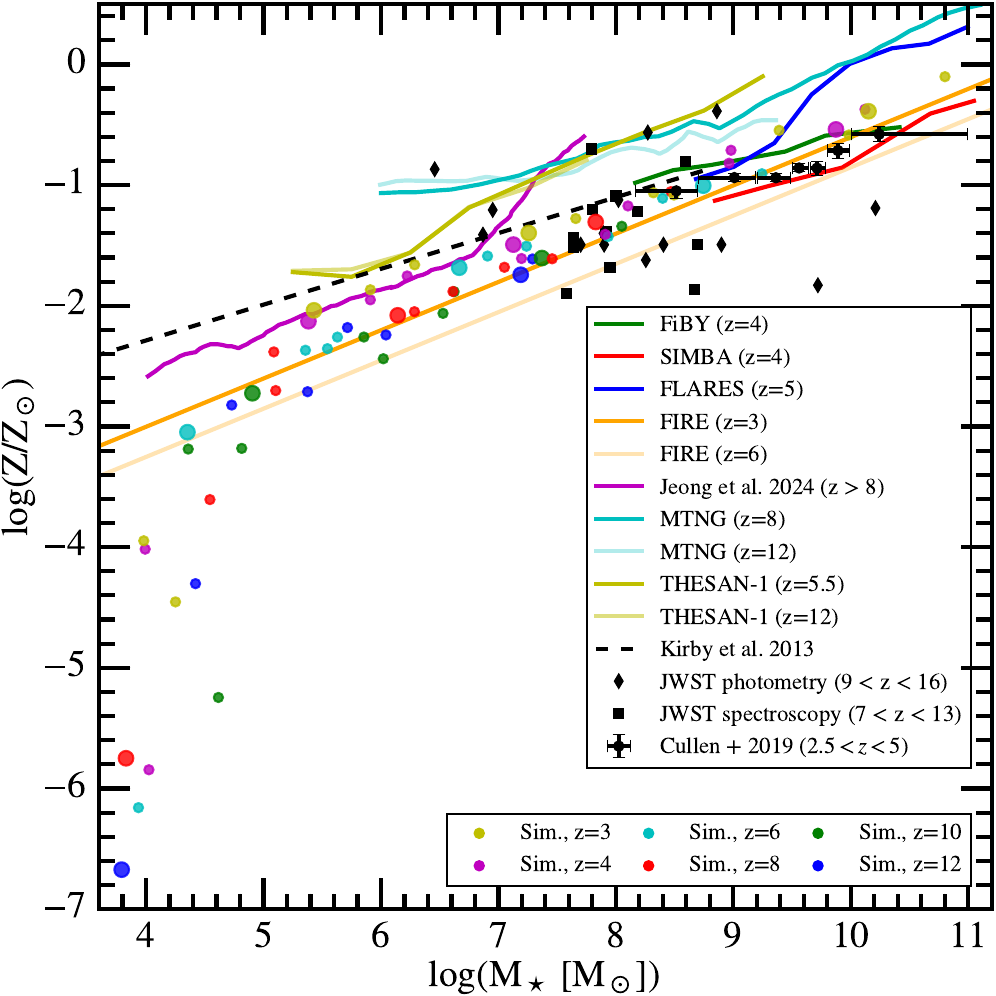}
    \caption{ The stellar mass (within twice the stellar half mass radius) - stellar metallicity (mass-weighted mean within the stellar half mass radius) relation for the central galaxies of all target haloes at different redshifts. We always choose the highest resolution run for each halo and additionally highlight the three simulations we focus on in the main part of the paper by increasing their symbol. We compare our results with the following cosmological simulations: {\small FiBY} \citep{fiby}, {\small SIMBA} \citep{dave2019simba}, {\small FLARES} \citep{wilkins2023first}, {\small FIRE} \citep{ma2016origin}, the simulation E001 from \protect\cite{Jeong2024}, MilleniumTNG \citep{pakmor2023millenniumtng, kannan2023millenniumtng}, and \thesanone \citep{Thesan1,ThesanDR}. 
    We also add observational results from the {\small VANDELS} survey \citep{cullen2019vandels}, photometry-only data from JWST \citep{Furtak2023, Robertson2023} and spectroscopically confirmed data \citep{Wang2023, Tacchella2023, Curti2024, Carniani2024, CurtisLake2023} as well as data from dwarf galaxies within the local group \citep{Kirby2013}.
    Our results compare well with {\small FiBY} and the {\small VANDELS} survey. 
    The observational results from JWST show a larger scatter than our results, though especially the spectroscopically confirmed objects agree on average with our simulation. However, they lie rather on the lower end of the metallicity. 
    This could be explained by the typically higher redshift of the observations compared to our data points at the same stellar mass. \protect\cite{Jeong2024} finds a higher metallicity than our results in their simulation, especially when considering their higher redshift. This could be explained by their different feedback model, especially their specific Pop III star formation model with higher metal yields.
We find consistently higher metallicities than FIRE for $M_\star / \text{M}_\odot > 10^5$. Nevertheless, the slope between {\small FIRE} and our simulations is very similar. }
    \label{fig:stellar_metallicity_stellar_mass_relation}
\end{figure*}

For each simulation we produced 189 Snapshot files that contain all properties of gas, dark matter and stellar particles with a time cadence of $10~\mathrm{Myrs}$, from $z=16$ 
down to $z=3$. We also generated halo catalogues using the friends-of-friends (FOF) algorithm \citep{Davis1985}, which only considers the spatial proximity of particles.
The FOF groups were further processed using the SUBFIND-HBT algorithm \citep{springel2001populating, Gadget4}, which identifies self-gravitating substructures known as subhaloes. 
We define the most massive subhalo as the central galaxy, while smaller ones are classified as satellite galaxies.\\
To differentiate between stellar populations, we use criteria based solely on the metallicity of stellar particles compared to the solar metallicity $\text{Z}_\odot = 0.0127$ \citep{wiersma2009chemical}:
Pop I stars have $Z / \text{Z}_\odot > 0.1$, Pop III stars are characterized by $Z / \text{Z}_\odot < 10^{-6}$,  and Pop II stars fall in between. 
As discussed in the introduction, the critical metallicity for the transition from Pop III to Pop II star formation is not well constrained. Previous studies have used $Z = 10^{-4}\,\text{Z}_\odot$ \citep{jaacks2019legacy, liu2020gravitational, venditti2023needle}, $Z = 10^{-5}\,\text{Z}_\odot$ \citep{Ricotti2016, sarmento2018following,sarmento2022effects,Brauer2024} or $Z = 5 \times 10^{-6}\,\text{Z}_\odot$ \citep{skinner2020cradles}.
We adopt a more conservative threshold due to the absence of an explicit Pop III star formation and feedback model in our simulations, particularly given the potential for higher metal yields from these stars \citep{takahashi2018stellar}.
Our selection criterion only identifies stars that form very close to the metallicity floor as Pop III stars. Therefore, the values we present in this paper should be viewed as lower limits.
As an example, we show the halo m10.8\_8x and its environment at redshift $z=7$ in \cref{fig:51_cosmo_new_bin}.
The central dark matter halo is less concentrated than the gas density, and we observe a plethora of dark matter substructures. 
Although the gas is enriched with metals on large scales, surviving metal-free gas still allows the formation of Pop III stars outside the target halo.

\section{The stellar metallicity distribution}
\label{sec:stellarMetallicityDistribution}
\begin{table}
    \centering
    \begin{tabular}{c|c|c|c}
     z    &  C & $Z_7$ & $\log(M_{\star, \text{max}} /\text{M}_\odot)$\\
     \hline
      3   & 0.35 & $-1.50$ & 10.8 \\
      4   & 0.37 & $-1.57$ & 10.1\\
      5   & 0.40 & $-1.65$ & 9.5\\
      6   & 0.40 & $-1.70$ & 9.2\\
      8 & 0.42 & $-1.71$ & 8.8\\
      10 & 0.48& $-1.77$ & 8.1\\
      12 &0.48 & $-1.75$ & 7.3 \\
      \hline
    \end{tabular}
    \caption{The parameters for the best linear fit as presented in Equation~(\ref{eq:bestFit}) to the stellar mass - stellar metallicity relation for $M_\star > 10^{5.3} \msun$ for different redshifts.
    We also show the maximum stellar mass we find in any single halo as an upper limit for the valid range for our fit.}
    \label{tab:bestFits}
\end{table}
We begin our investigation by showing in \cref{fig:stellar_metallicity_stellar_mass_relation} the stellar mass - stellar metallicity relation for all central subhaloes within the target haloes at different redshifts.
The stellar mass is measured within twice the stellar half-mass radius, and the stellar metallicity is defined as the mass-weighted mean metallicity of all star particles within the stellar half-mass radius.
 All haloes with stellar masses $M_\star$ above $10^{5.3} \msun$ follow a relatively tight relation. However, for smaller masses, we observe a significant scatter down to the imposed metallicity floor, particularly at higher redshifts. 
The slope of the relation remains consistent over time; however, due to previous stellar feedback, the entire relation shifts to higher metallicities over time. By performing a least squares fit to the equation:
\begin{equation}
    \log(Z/\text{Z}_\odot) = C  \log_{10}(M_\star/\text{M}_\odot - 7) + Z_7 \, ,
    \label{eq:bestFit}
\end{equation}
we find the values presented in \cref{tab:bestFits} for the slope $C$ and the offset $Z_7$.
We only included the central galaxies of the target haloes with $M_\star > 10^{5.3} \msun$.
The slope decreases slightly over time from $0.42$ at $z=8$ to $0.35$ at $z=3$. These values are similar to those found in \cite{ma2016origin} for the {\small FIRE} suite, which used a redshift-independent slope of $0.4$ for general fits. 
According to their Figure 4, the slope for $z=3$ is slightly smaller, and for $z=6$ , it is slightly larger than this fiducial value.
The offset $Z_7$ generally increases over time.
It is important to note that we used a very limited number of fitting points, particularly at high redshift. We could enhance the sampling size by including satellite galaxies and additional haloes within the high-resolution regions. However, this could introduce a bias, as satellite galaxies can have systematically different stellar mass - stellar metallicity relations \citep[e.g.][]{pasquali2010ages,gallazzi2021galaxy}.

Our results compare well with observational constraints from the VANDELS survey \citep{cullen2019vandels} and individual measurements with JWST. 
The latter shows, in general, a larger scatter than we observe in our simulations. For $M_\star > 10^6\msun$, our results are also compatible with local measurements of dwarf galaxies within the local group \citep{Kirby2013}, which is expected if they form the majority of their stars during the EoR.  
At the high mass end we predict similar values as those from the {\small FiBY-XL} simulation \citep[we use the values presented in][]{cullen2019vandels}.
The {\small SIMBA} \citep{dave2019simba} simulation shows slightly lower metallicities \citep[we use again the values from][]{cullen2019vandels} while {\small FLARES} \citep{wilkins2023first} overpredicts the metallicity. 
As discussed in \cite{wilkins2023first}, this discrepancy can be partially attributed to an alpha enhancement of stars in the early universe. 
The Millenium TNG simulation \citep{pakmor2023millenniumtng,kannan2023millenniumtng} and the original \thesanone simulation, show significantly higher stellar metallicities with minimal redshift evolution.
We also compared our results to those of the {\small FIRE} suite \citep{ma2016origin}, which explored a much larger parameter space than the other studies using zoom-in simulations. After rescaling their results from  $\text{Z}_\odot=0.02$ to $\text{Z}_\odot=0.0127$, we still find around 0.4dex higher metallicities in our simulations at $z=3$. 
\cite{Jeong2024} reported significantly larger metallicities in their zoom-in simulations, particularly at high redshifts. This could be explained by their larger yields stemming from an explicit Pop III stellar evolution model.
Although we only show the result from one of their simulations, the slope remains similar to our results up to a stellar mass of $ 10^7\msun $.

\begin{figure*}
    \centering
    \includegraphics[width=1\linewidth]{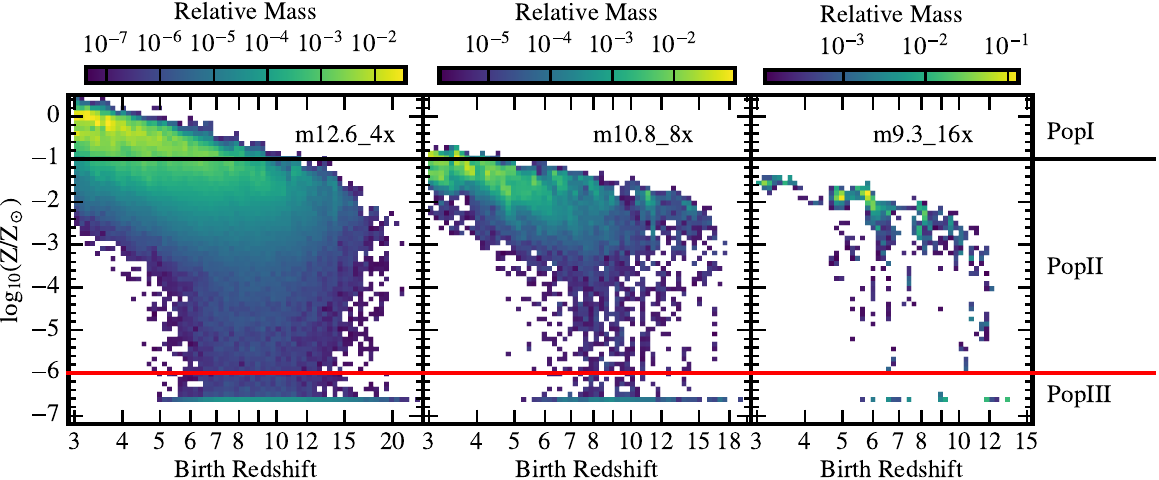}
    \caption{We show a histogram of the birth mass of all stellar particles that can be found in three high-resolution haloes at redshift $z=3$ as a function of their birth redshift and metallicity normalized by the solar metallicity.
 All haloes contain Pop III stars ($\rm Z < 10^{-6}\,\text{Z}_\odot$) born at the end of the epoch of reionization ($z \approx 5 - 5.5$), which mostly form in satellite galaxies or ex-situ. On average, metal-rich stars can be found in larger haloes and are born at lower redshift. \cref{fig:age_metallicity_stars} contains the data for the remaining 11 target haloes.}
    \label{fig:age_metallicity_stars_3}
\end{figure*}

\begin{figure*}
    \centering
    \includegraphics[width=1\linewidth]{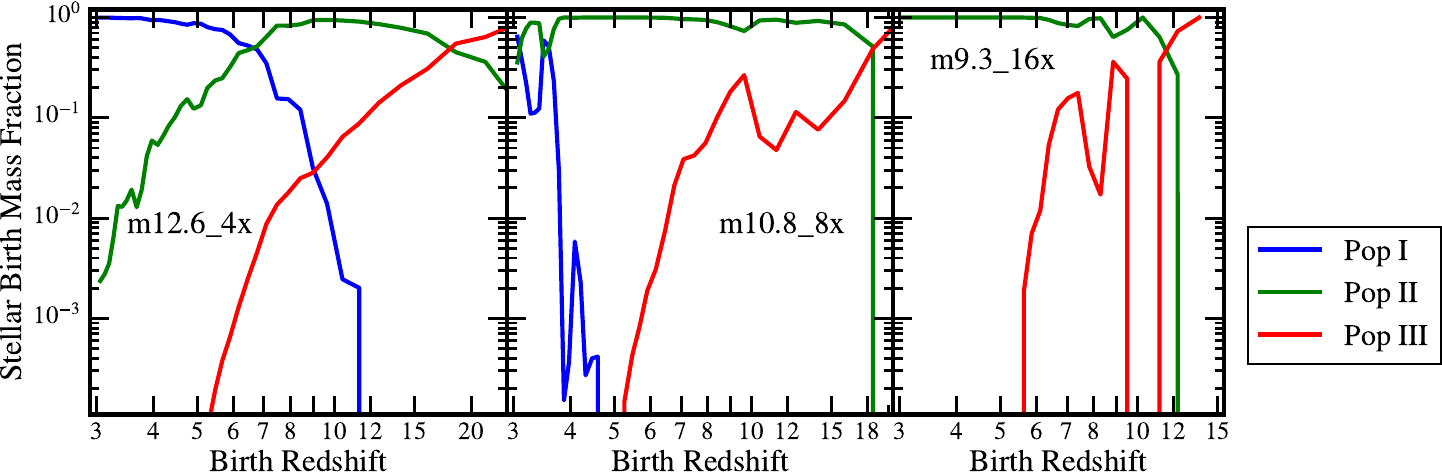}
    \caption{The birth mass fraction of different stellar populations as a function of the birth redshift. We averaged the results over $100\,\rm Myr$ and used the birth mass of the star particles. We only consider stars in the target halo at $z=3$, but they can be born outside of it.
    Pop III star formation dominates at high redshift, and only in the most massive halo do we find a transition to Pop I star formation before redshift $z=3$. 
    Nevertheless, we can find Pop III stars formed at the end of the EoR at $z = 5-6$ in all haloes.
    \cref{fig:ratio_star_formation_populations_per_halo} contains the data for the remaining 11 target haloes.}
    \label{fig:ratio_star_formation_populations_per_halo_3}
\end{figure*}

As a next step, we present in \cref{fig:age_metallicity_stars_3} the metallicity and birth redshift distribution of all stars found in the target haloes at redshift $z=3$. 
The more massive haloes show a continuous evolution of star formation and stellar metallicity, whereas the smallest one shows artefacts of a more bursty star formation history (see the companion paper \cite{zoomJacob} for a more detailed discussion of the star formation history).
On average, younger stars tend to be more metal-rich, and the maximum metallicity increases with halo mass.
This can be attributed to the deeper gravitational potential of more massive haloes, which reduces the efficiency of supernovae blowing enriched gas out of the halo, ultimately leading to a higher star formation efficiency \citep[e.g.][]{kitayama2005supernova, whalen2008destruction}.
In our simulations, all target haloes contain a population of extremely metal-poor stars that can form until redshift $z=5$, shortly after the end of reionization.
Following this period, the minimum metallicity rapidly increases to $\approx 10^{-4}\,\text{Z}_\odot$, which is the maximum metallicity used in other simulations to define Pop III stars.
The minimum redshift for Pop III star formation would move to $z\approx 4.5$ for this less strict criterion.
We note that the value of $10^{-4}\,\text{Z}_\odot$ aligns with the self-consistently established metallicity floor found in simulations that directly model Pop III stars \citep[e.g.][]{Brauer2024}. 
Most other studies analyzing Pop III star formation at lower redshift ended before the end of reionization,  e.g. at $z=7$ \citep{xu2016late,jaacks2019legacy} or $z=6$ \citep{johnson2013first} and were unable to track the end of Pop III star formation. 
A notable exception is \cite{liu2020gravitational}, which found a continuous but decreasing Pop III star formation until $z=4$. However, this study's lack of radiative transfer does not enable it to self-consistently model reionization, which may explain the absence of a significant decline in the Pop III star formation rate density (SFRD) while reionization progresses.

\begin{figure*}
    \centering
    \includegraphics[width=1\linewidth]{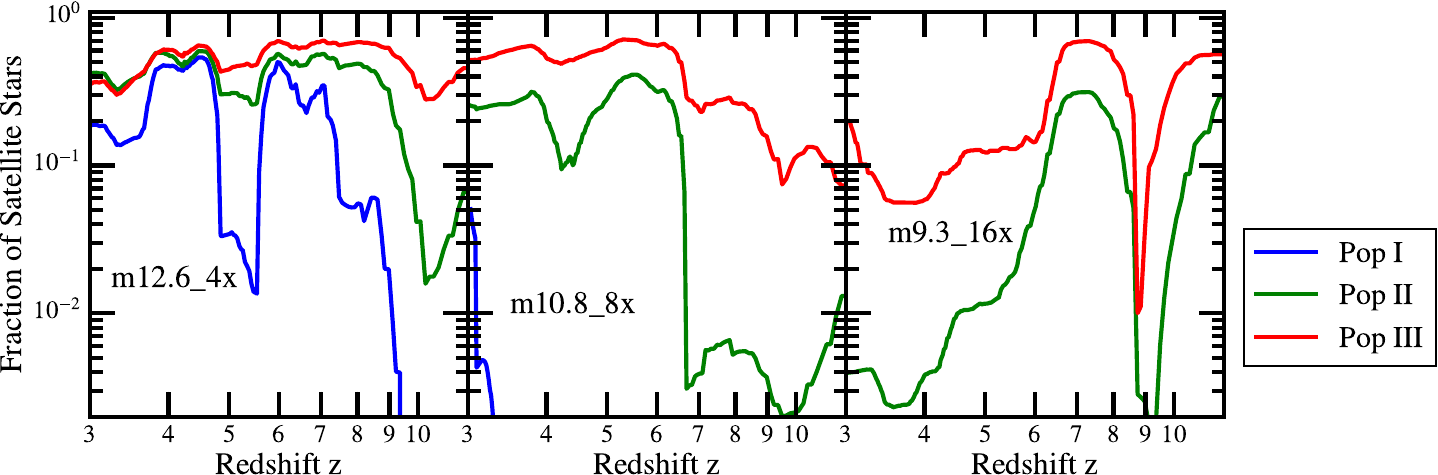}
    \caption{The fraction of different stellar populations (represented by three different solid lines) found in satellites of the target halo at different times averaged over $100\, \rm Myr$. Only a neglectable fraction of stars within the halo is not bound to any subhalo.
    Pop I stars can be mostly found in the central galaxy, while Pop III stars are found more often in satellites. The satellite fraction can significantly increase during major mergers when the halo finder identifies the two merging haloes for the first time as a single FOF group.
    If the haloes do not directly merge the satellite fraction can also drop again when the halo finder identifies the two separate haloes again. This can be seen e.g. for Pop I stars for m12.6\_4x.
    We only consider star particles that are at redshift $z$ in the target halo and use the initial birth mass of stars to calculate mass fractions to reduce the impact of a Pop III IMF.
    }
    \label{fig:individual_evolution_stars_per_position_3}
\end{figure*}
For a more quantitative analysis, we present the birth mass fraction of each stellar population as a function of age in \cref{fig:ratio_star_formation_populations_per_halo_3}.
More massive haloes typically exist in over-dense regions with earlier structure formation, which leads to an earlier onset of star formation \citep{Regan2023}. This trend can also be seen in our simulations.
Star formation begins around redshift $z\approx 24$ ($\rm 136\, Myr$), which is slightly later than the canonical value of $z\approx 30$ for the start of Pop III star formation \citep{klessen2023first}.
This discrepancy can be explained by our lower resolution, which is used to simulate the most massive target haloes and does not allow us to fully resolve the earliest minihaloes capable of forming stars.
Feedback from the first stars pollutes their birth halo with metals and leads to the formation of the first enriched Pop II stars at $z=21$ ($\rm 168\, Myr$), similar to $z\approx 22$ as found in \cite{Brauer2024}.
Pop II star formation starts to dominate in most target haloes around redshift 12 to 18, with an earlier transition in the more massive haloes due to the higher star formation rates.
In the more massive halo, we also observe a transition to 
Pop I star formation at $z\approx 6 -7$.
Nonetheless, in all three zoom-in regions, we find the formation of Pop III stars down to redshift $5$. 
The timing of the transition to Pop II dominated star formation is consistent with small box simulations \citep[e.g. $z\approx 17$,][]{sarmento2018following, Brauer2024}, large box simulations \citep[e.g. $z\approx 13$,][]{venditti2023needle} and semi-analytic models \citep[e.g. $z\approx 15$,][]{Hartwig2024}. However, the exact timing also depends on the critical transition metallicity.
To quantitatively compare with previous studies, we would need global star formation rate densities for different stellar populations.
In our case, these can be computed in the high-resolution regions; however, those regions are biased and would exhibit a similar spread in transition times as the target haloes.
Estimates for global star formation rates can be obtained by a detailed comparison with the same regions in the parent box, as \cite{zoomIntro} did for other properties, such as the galaxy stellar mass function or the UV luminosity function. This analysis will be included in an upcoming paper.

By tracing the target halo back in time, we calculate the number of stars found in its satellites as a function of redshift.
For each snapshot, we classify the location of each star particle within the target halo, more precisely whether it can be found in the central subhalo, in a smaller subhalo (satellite), or it is not bound to any subhalo. 
The latter case involves only a negligible number of particles.
To calculate the satellite fraction for stars, we use their initial birth mass rather than their current mass.
While this approach neglects mass loss from stellar feedback, particularly from supernovae, it reduces the dependence of our results on assumptions about the IMF for Pop III stars.
The satellite fraction can significantly increase during mergers before the two main subhaloes merge since the central galaxy of the smaller halo is identified as a satellite.
This can be seen in \cref{fig:individual_evolution_stars_per_position_3} for the simulation m12.6\_4x at e.g. $z \approx 5$.
Pop I star formation requires a sufficiently massive halo, as we have seen before in \cref{fig:ratio_star_formation_populations_per_halo_3}. Consequently, only during major mergers can a significant number of Pop I stars be found in satellites.
Pop II stars can form in smaller haloes and, therefore, can also end up in satellite galaxies during minor mergers, as e.g. seen in m12.6\_4x in \cref{fig:individual_evolution_stars_per_position_3}. 
These stars are the primary component of the smaller haloes and are generally located in the central part of those haloes. 
Pop III stars only form in the smallest haloes and are therefore more likely to be found in satellites.
A closer inspection of the radial mass profiles shows that they are less concentrated than the other populations, with young Pop III stars often located on the outskirts of galaxies. 
If any Pop III stars with masses below $0.8 \msun$ survive to the present time, they are likely to be predominantly found in dwarf satellite galaxies \citep[consistent with e.g.][]{magg2018predicting} or within the galactic halo \citep{Hartwig2015}.

\section{How do metal-poor stars form?}
As discussed in the previous section, pristine gas can form Pop III stars until the end of reionization.
In this section, we will analyze the evolution of pristine gas around the target haloes to better understand the formation mechanisms and birth sites of these stars. Additionally, we will trace the Pop III stars back to their birth haloes.

\label{sec:PopIIIStarFormation}
\subsection{The evolution of the gas phase}
\begin{figure*}
    \centering
    \includegraphics[width=1\linewidth]{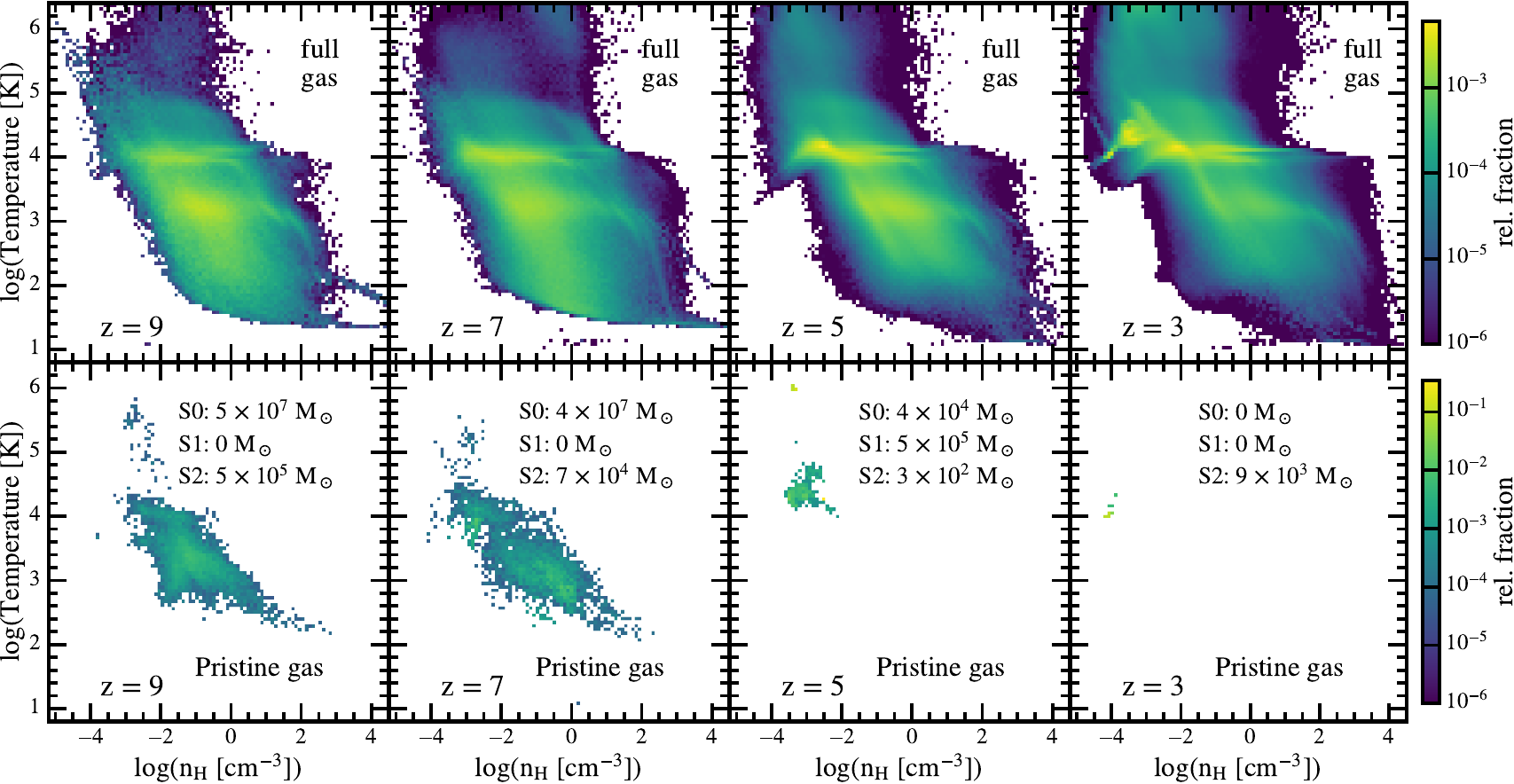}
    \caption{Phase diagram of gas in the three target haloes at four different redshifts. The first row shows all gas while the second one only shows pristine gas ($Z < 10^{-6}\,\text{Z}_\odot$).
    We averaged the relative fractions of all three haloes in each bin and introduced the abbreviations S0 (m12.6\_4x), S1 (m10.8\_8x) and S2 (m9.3\_16x) for easier display. In the second row we state the total mass of pristine gas within each halo at the specific redshift.
    The full gas shows a complex structure with a concentration of gas at $\rm T\approx 10^4\,K$, the temperature at which atomic cooling becomes efficient. 
    The pristine gas is confined to a small region in the phase space that corresponds to recently accreted gas and is almost fully removed at $z=3$. Only for $z> 5$, we find star-forming, pristine gas. 
    }
    \label{fig:phase_diagram_metallicity_3}
\end{figure*}

We start our analysis of the gas phase by calculating phase space diagrams for the target haloes at different redshifts. 
We show in \cref{fig:phase_diagram_metallicity_3} the averaged diagrams for the full halo gas and the pristine gas. 
The full gas within the target haloes shows a complex pattern, spanning five orders of magnitude in temperature and eight in density. 
Especially at lower redshift, we observe a concentration of gas at $T \approx 10^4~\mathrm{K}$, which corresponds to the temperature above which atomic hydrogen cooling becomes efficient. 
We also find cool, low-density gas ($T < 10^3$\,K, $n_\text{H} < 0.1\,\mathrm{cm}^{-3}$) that is shielded by denser gas;  however, the fraction of this gas substantially decreases after $z=7$. 
Without self-consistent radiative transfer but with a uniform UV background, this gas would almost instantaneously become ionized and heated (Zier et al. 2025).
Almost no gas can be found below the
 redshift-dependent temperature of the CMB given by $T_{\rm CMB}(z) = \left(1+z\right) T_{\rm CMB,0}$,  where the present temperature is $\rm T_{\rm CMB,0} = 2.726\,K$ \citep{fixsen2009temperature}.
 We note that all simulations use a temperature floor of $12\,$K, which becomes relevant at lower redshift.
Within the haloes, we also find relatively dense and hot gas ($T> 10^{4.5}$\,K), which has recently been heated by supernovae or structure formation shocks. The low metallicity gas that has the potential to form Pop III stars occupies only a small portion of the phase diagram and adheres to a relatively strict relationship between temperature and density. Our conservative metallicity criterion for Pop III star formation explains this, as any interaction with metal-enriched stellar ejecta from supernovae or stellar winds would move the gas metallicity above this condition.
Consequently, the only available heating mechanisms are associated with the accretion shock and the interaction with the radiation field.
Even the densest low metallicity gas reaches only a minimum temperature of $200$\,K, which corresponds to the minimum temperature achievable through molecular hydrogen cooling \citep{greif2015numerical}.
We find star-forming, low-metallicity gas only at $z=7$ and $z=9$. Further analysis reveals that this gas is mostly found in subhaloes (satellite galaxies) rather than in the central galaxy.
\begin{figure*}
    \centering
    \includegraphics[width=1\linewidth]{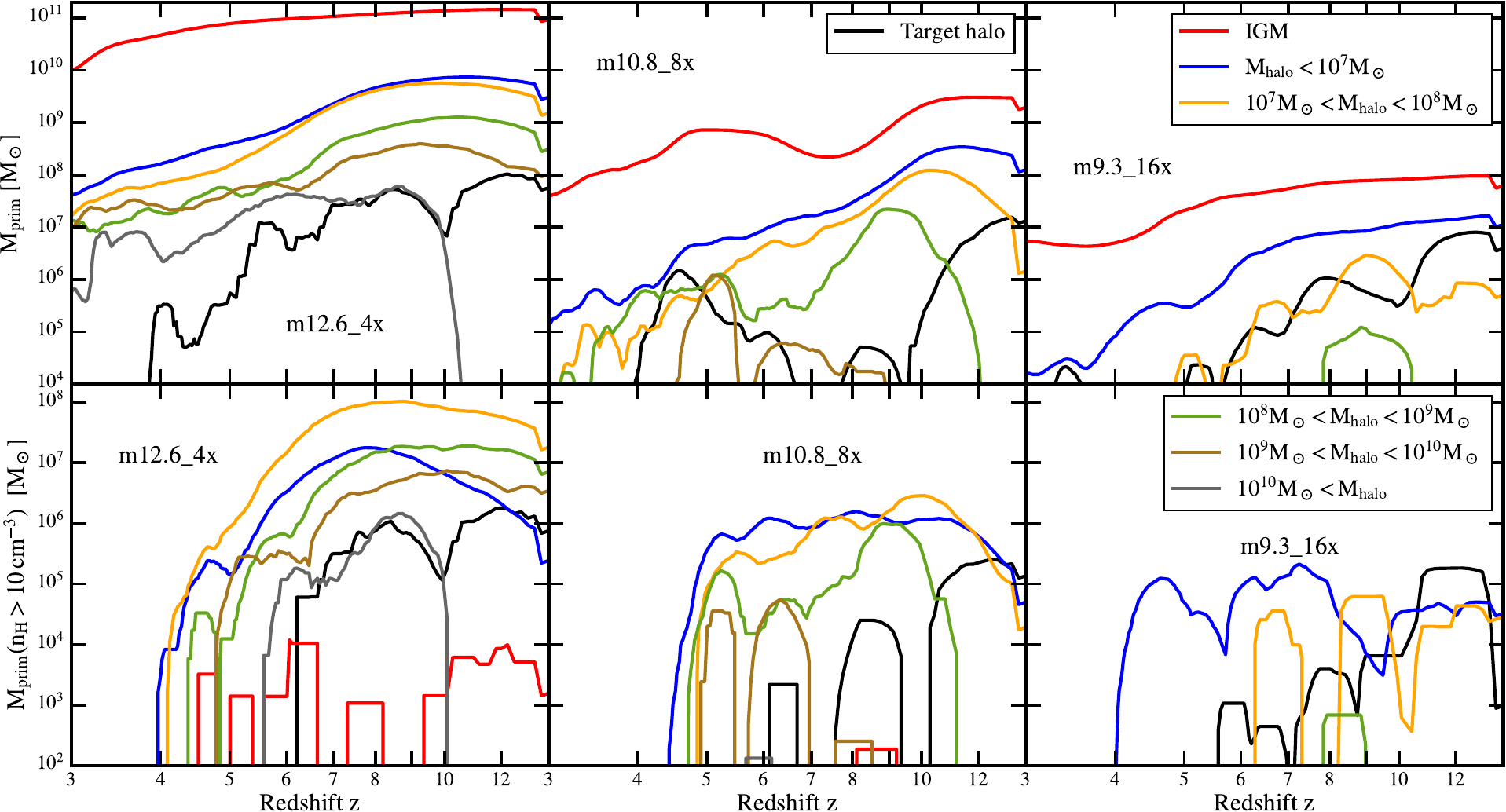}
    \caption{The total amount of primordial gas able to form Pop III stars ($Z < 10^{-6}\,\text{Z}_\odot$) that can be found within a fixed comoving distance of $3\times R_{200}(z=3)$ around the target halo. 
    We show all the gas (top) and only the potentially star-forming, high-density one (bottom). The black line represents the gas that can be found in the target halo, while the red line represents the IGM, which is defined as all gas not being part of any halo. 
    The other coloured lines represent the gas that can be found in other high-resolution haloes around the target halo. We split this gas according to the mass of the current halo they are associated with, and different colours represent different halo mass ranges.
     At all times, most of the primordial gas can be found in the IGM, while the star-forming one can be mostly found in haloes with $M_{\rm halo} < 10^8\msun$. The results are averaged over $100\,\rm Myr$. }
    \label{fig:pristine_gas_data}
\end{figure*}

To better understand the birthplaces of Pop III stars, we analyse the high-resolution gas outside of the target halo, specifically within a comoving distance of three times the virial radius of the target halo at redshift $z=3$. 
For each low metallicity gas cell ($Z< 10^{-6}\,\text{Z}_\odot$), we check the mass of its parent halo and define the intergalactic medium (IGM) as all gas outside of any halo.
For the gas that is not part of the IGM and the target halo, we introduce several bins based on the mass of their parent halo.
In \cref{fig:pristine_gas_data}, we show the temporal evolution of the mass of all low-metallicity gas, as well as the mass of high-density low-metallicity gas, which we define by using the same density threshold $n_\mathrm{H} = 10~\mathrm{cm}^{-3}$ as used for star formation.
This gas represents a superset of the gas which is allowed to form stars since the latter one is additionally Jeans unstable.
Most of the low-metallicity gas in all simulations can be found in the IGM and low-mass haloes with masses below $10^7\msun$. This gas continues to exist until the end of our simulations at redshift $z=3$, but its amount decreases over time due to metal enrichment. 
Nevertheless, metal mixing in the IGM seems to be inefficient.
The presence of large reservoirs of pristine gas in minihaloes and the IGM is consistent with previous studies; e.g. \cite{Wise2012} found almost pristine haloes below $M_{\rm halo} \approx 10^{7.3} \msun$ at $z=7$
and \cite{liu2020did} found a volume filling fraction of only 2\% at $z=4$ of enriched gas with $Z > 10^{-4}\,\text{Z}_\odot$ in a cosmological box.
However, detailed quantitative comparisons would require full cosmological boxes, as the regions around the target haloes can be biased. 
In our simulations, the pristine gas in the IGM does not reach the densities necessary for star formation; instead, Pop III star formation is dominated by haloes below $10^9\msun$.
In haloes with masses below $10^7\msun$, a smaller fraction of low-metallicity gas reaches higher densities, which can be explained by the lower efficiency of molecular hydrogen cooling in very small haloes \citep{glover2012first}.
\begin{figure*}
    \centering
    \includegraphics[width=1\linewidth]{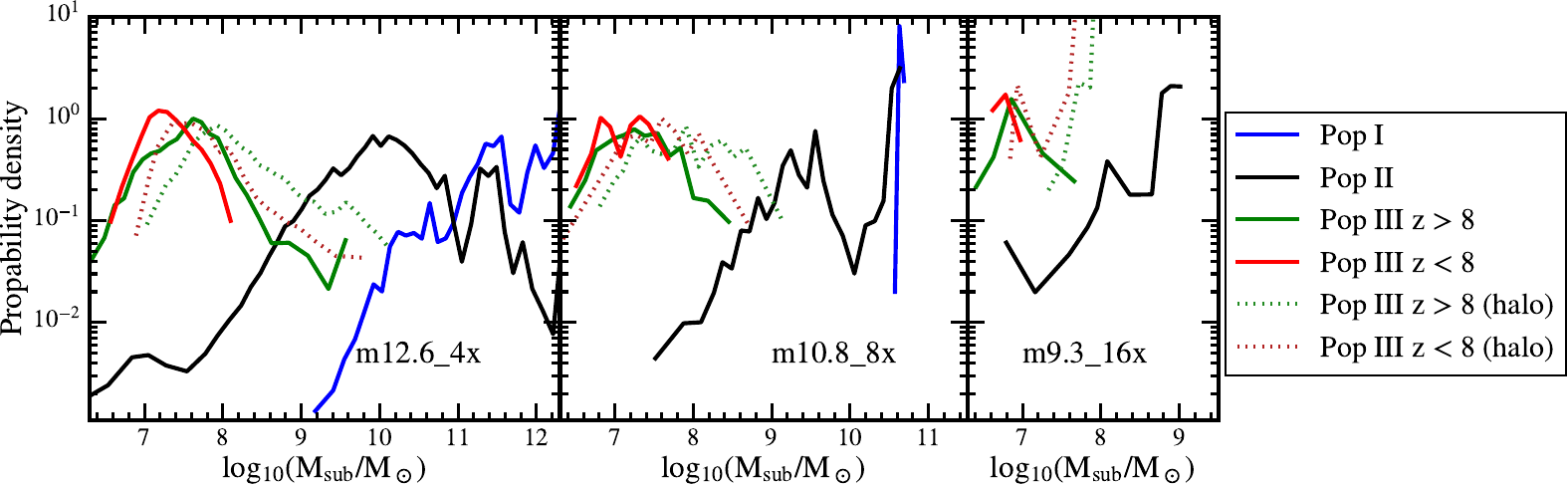}
    \caption{The mass distribution of the subhaloes in which different stellar populations are born. We use all stars found at redshift $z=3$ in the target haloes and the data is normalized so that integral over the full subhalo mass range yields 1.
    We divided Pop III stars into those born before (green line) and after $z=8$ (red lines) and show additionally their birth halo mass distribution (dotted lines). 
    Pop I and Pop II stars mostly follow their host halo mass distribution with time, while Pop III stars are almost exclusively formed in subhaloes with masses below $10^8\msun$. 
The data for the remaining 11 target haloes can be found in \cref{fig:birth_subhalo_distribution_full}.}
    \label{fig:birth_subhalo_distribution_3}
\end{figure*}

\begin{figure*}
    \centering
    \includegraphics[width=1\linewidth]{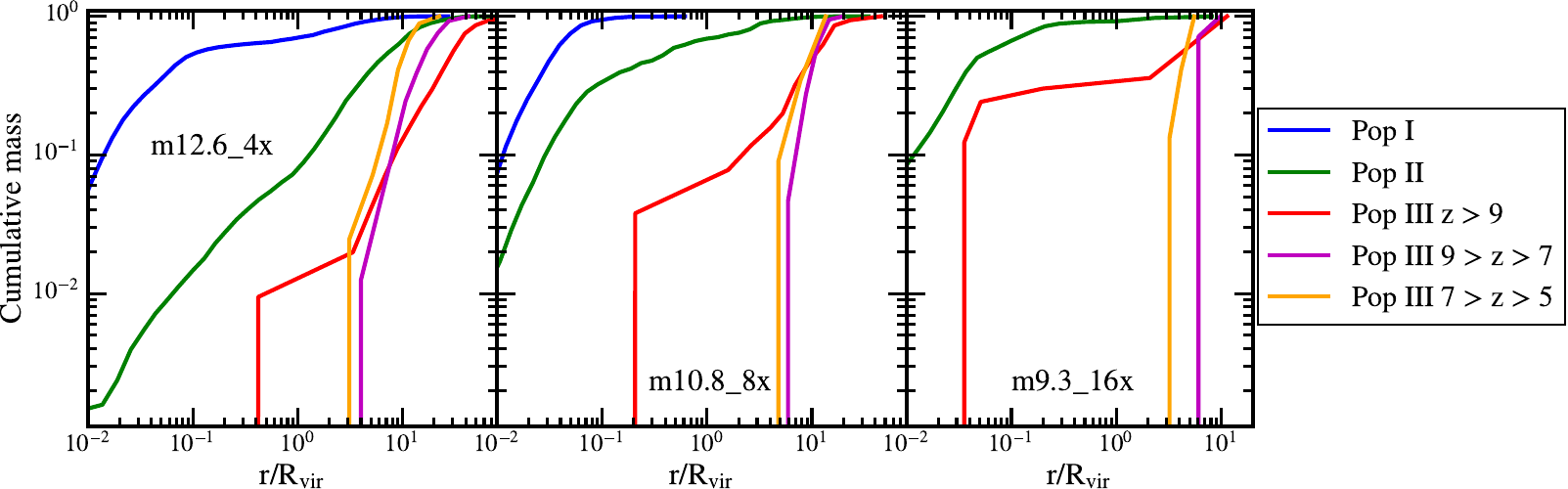}
    \caption{Cumulative mass distribution as a function of distance to the target halo centre at the redshift of birth for three different stellar populations.
    We split Pop III stars additionally into three redshift bins depending on the time of their formation.
    The distances are normalized by the virial radius of the target halo at the redshift of their birth time.
    Except for the largest halo that grows through mergers, Pop I (blue line) and Pop II (green line) stars mostly form within the virial radius, while Pop III stars after $z=9$ (magenta and orange line) form almost exclusively outside the virial radius. The data for the remaining 11 target haloes can be found in \cref{fig:radial_mass_distribution_origin_cum_rel_r_vir}.}
    \label{fig:radial_mass_distribution_origin_cum_3}
\end{figure*}

\subsection{Origin of Pop III stars in the target halo}
To identify the birth sites of Pop III stars, we trace all stars found in the target haloes at redshift $z=3$ back to the first snapshot after they formed.
By neglecting potential mergers that may occur between the time of the snapshot and the actual star formation, we can effectively determine each star's birth halo and subhalo.
This approximation is justified by the high temporal resolution of the snapshot files, which are taken approximately every 10\,Myr. 
In \cref{fig:birth_subhalo_distribution_3}, we show the mass distribution of subhaloes in which the three stellar populations formed.
Generally, Pop I stars form in more massive subhaloes than Pop II stars, which is a direct consequence of the first one's later formation time and their preferred formation in the central galaxy. 
Efficient Pop I star formation requires a minimum subhalo mass of $10^{10}\msun - 10^{11}\msun$.
In comparison, Pop II stars show a broader mass distribution and can even form in enriched minihaloes, although their peak formation occurs at $\approx 10^{10}\msun$.
Pop III stars behave quite differently, forming in subhaloes with typical masses ranging from $10^{6.5}\msun$ to $10^8\msun$.
The lower mass limit, which is smaller in the better-resolved m9.3\_16x simulation, shows some resolution dependency.
This behaviour is expected, as in the m12.6\_4x simulation, the smallest star-forming haloes are only resolved by a few hundred DM particles.
The exact lower boundary for star formation is a function of the local Lyman-Werner radiation background, which varies with redshift \citep[e.g.][]{greif2006two}, and the relative streaming velocity between DM and baryons.
Without radiation, the minimum estimates are $\approx 10^6 \msun$ \citep{fialkov201321, kulkarni2021critical, klessen2023first} at $z=10$. However, strong radiation can raise this limit to $10^7 \msun$ \citep{o2008population}.
Thus, our lower limit is consistent with other studies and analytical estimates \citep{glover2012first}. 
The upper limit for Pop III star formation below $z=8$ coincides with the transition to atomic cooling haloes, occurring at masses between $5\times 10^7\msun$ and $10^8 \msun$ at $z < 10$ \citep[see equations in ][]{hummel2012source,hartwig2022public}. These haloes are more efficient at forming stars due to additional cooling provided by atomic hydrogen.
They may also be sites for the formation of massive black hole seeds through the direct collapse of pristine gas, which is not incorporated into our model. 
The birth halo mass distribution of Pop III stars closely resembles the subhalo distribution, although the peak is slightly shifted to larger masses.  This shift occurs primarily due to the inclusion of smaller subhaloes and unbound material in the halo mass calculation.
A tail also extends to halo masses of $10^{10}\msun$ even for the lower redshift bin, which is caused by Pop III star formation in satellite galaxies.

We also calculated the distribution of distances to the centre of the target halo at the formation time of stellar particles and normalised it with the virial radius at this time.
We use all stellar particles found at $z=3$ in the target halo and present the results in \cref{fig:radial_mass_distribution_origin_cum_3}.
Pop I stars tend to form near the centre of the target halo, while the concentration of Pop II stars varies with halo mass. 
In smaller haloes, which are Pop II dominated even at lower redshift, they form more concentrated. However, in the most massive halo, Pop II stars predominantly form outside the virial radius.
In this situation, several progenitors of the target halo can form Pop II stars, leading to a higher fraction of this stellar population forming ex-situ. 
 Pop III stars below redshift $z = 9$ exclusively form outside the virial radius and are later accreted by the target halo.
 This offset is also observable in the visualization \cref{fig:51_cosmo_new_bin} and as we discuss in \cref{subsec:ObservationalStudies}, it is also consistent with potential observations of Pop III dominated systems at lower redshift \citep{vanzella2023extremely, wang2024strong,maiolino2024jades}.

\section{Discussion}
\label{sec:discussion}
\subsection{Comparison with theoretical studies}
Theoretical studies of Pop III star formation can be categorized into two main categories. The first category focuses on the detailed formation processes of these stars during the period when they dominated the cosmic star formation rate (typically at $z> 11$).
 These studies examine the internal properties of minihaloes and the conditions necessary for star formation  \citep[e.g.][]{abel2002formation, Yoshida2003, o2007population, smith2015first, kulkarni2021critical, liu2022effects,kiyuna2023first,lenoble2024simulations, correa2024role,smith2024does}. Some studies in this category also attempt to resolve an IMF using a zoom-in approach \citep{hirano2015primordial,stacy2016building}.
The second category of studies aims to understand the statistical properties of Pop III star formation using cosmological boxes.
Our study falls into the second category, but we use a zoom-in approach that allows for higher resolution, although this choice prevents us from directly obtaining the cosmic-averaged star formation rate density. 

\cite{venditti2023needle} analyzed eight cosmological simulations from \cite{di2023assembly}, which were performed using the dustyGadget code in a volume of $50 \rm \mathit{h}^{-1} \cMpc$  with a DM/gas mass of $3.53 \times \rm 10^7 \mathit{h}^{-1} \msun$ /$5.56 \times 10^6 \rm \mathit{h}^{-1} \msun$ per particle. These simulations utilized an effective equation of state model \citep{springel2003cosmological} and incorporated a cosmological dust model but did not include radiative transfer.
They found continuous Pop III star formation until the end of the EoR ($z\approx 6-8$), typically occurring at the outskirts of metal-enriched regions or isolated pristine gas clouds.

\cite{jaacks2018baseline,jaacks2019legacy} performed simulations using the GIZMO code in a box with a side length of $4 \rm \mathit{h}^{-1} cMpc$ and DM/gas-particle mass $4.31 \times 10^4 \rm \mathit{h}^{-1} \msun$ / $9.64 \times 10^3 \rm \mathit{h}^{-1} \msun$ until $z=7.5$. They did not include radiative transfer but used different feedback models for Pop III and Pop II stars. They found continuous Pop III star formation until the end of their simulation, mainly dominating in haloes with total masses less than $10^9 \msun$.
\cite{liu2020gravitational,liu2020did} performed similar simulations using GIZMO  in a box with a side length of $4 \rm \mathit{h}^{-1} cMpc$ and DM/gas-particle mass $5.2 \times 10^4 \rm \mathit{h}^{-1} \msun$ / $9.4 \times 10^3 \mathit{h}^{-1} \msun$  until $z=4$. They found continuous Pop III star formation until the end of their simulation, with more massive haloes dominating at this later redshift with $M_{\rm halo}\approx 10^{10} \msun$.
Pop III stars at later times tend to form preferentially at the outskirts of these haloes. Discrepancies between these simulations and our simulations could be attributed to our more accurate treatment of reionization and the self-consistent formation of supernova-driven winds in our simulations compared to their model based on \cite{springel2003cosmological}.
The latter point can lead to more efficient metal mixing within haloes.

\cite{yajima2022forever22, yajima2023forever22} performed a zoom-in simulation of a region of size $3 \rm \mathit{h}^{-1} cMpc$ with gas particle resolution of $7.9 \times  10^3 \msun$ until $z=9.5$.
They found that galaxies with lower stellar masses tend to have a higher fraction of Pop III stars. Galaxies with stellar masses $ M_\star< 10^5 \msun$ are found to contain a significant fraction of Pop III stars (more than 10\%) at redshift $z=10$. Additionally, in larger haloes, regions of recent Pop III star formation are observed to be spatially separated from Pop II dominated regions and typically occur in recently accreted minihaloes.

The Renaissance project \citep{o2015probing} performed zoom-in simulations using the ENZO code in regions of varying cosmic densities.
\cite{xu2016late} specifically analyzed the void simulation with a volume of $220.5\rm\,cMpc^3$, which used a DM particle mass of $2.9\times 10^4\msun$ and was evolved until $z=7.6$.
They found continuous Pop III star formation in haloes with mass between $5\times 10^7 \msun$ and $10^8 \msun$. 
In smaller haloes, star formation was suppressed due to Lyman-Werner radiation from nearby neighbours,  while metal enrichment was ineffective in polluting the cores of these smaller haloes.

The THESAN-HR project \citep{borrow2023thesan} performed a simulation in a box of side length $4 \cMpc$ with DM/gas particle mass of $\rm 6.03\times 10^4\mathit{h}^{-1} \msun$/ $\rm 1.13\times 10^4\mathit{h}^{-1} \msun$ using the original \thesan setup.
They also found Pop III star formation until the end of reionization, though they used the ISM model from \cite{springel2003cosmological} and no Lyman-Werner radiation transport.
\cite{pakmor2022formation} reported Pop III star formation extending even down to $z=0$ in the IllustrisTNG simulations, although they did not resolve the multiphase ISM structure and relied on the model from \cite{springel2003cosmological}.

Overall, all these previous works predict low Pop III star formation rates at lower redshift, typically occurring in smaller haloes or at the outskirts of larger haloes, particularly in recently accreted minihaloes. These general trends agree with our results.
The exact end point of Pop III star formation is unclear, and it likely depends on the exact implementation of reionization, stellar feedback, and metal mixing. 
In our study, the fraction of dense, primordial gas significantly drops during reionization, which implies that Pop III star formation also ends around the end of the EoR.
Our high-resolution, explicit multiphase ISM model with self-consistent outflows, dust model, radiative transfer including Lyman-Werner radiation, and different cosmological environments make our simulation one of the most advanced larger-scale Pop III star formation simulations.
Nevertheless, some potentially important physical effects are missing, as discussed in \cref{subsec:caveats}.

\subsection{Comparison with observational studies}
\label{subsec:ObservationalStudies}
The lack of a clear observation of Pop III stars significantly complicates the comparison of existing observations with our simulations.
However, there are observational indications from JWST that could be attributed to Pop III stars or their direct descendants. Our study does not allow for the prediction of the physical properties of individual Pop III stars. Instead, it suggests the ongoing formation of Pop III stars until $z=5$, particularly in haloes with masses $M_{\rm halo} <10^{10} \msun$.  

\cite{welch2022jwst} discovered a highly lensed star at a redshift of $z=6\pm 0.2$ \citep{vanzella2023jwst}. This star resides in a galaxy with a stellar mass of $M_\star \approx 3\times 10^7\msun$, suggesting a lower mass limit for the halo of $10^9\msun$ \citep{schauer2022probability}.
While this finding is consistent with our simulations, the probability of it being a Pop III star remains relatively low \citep{schauer2022probability}. 

Another important indicator for Pop III stars are metal-poor but strong HeII-emission systems, which depend on the higher expected surface temperature of Pop III stars compared to Pop II stars \citep{venditti2024hide}. \cite{vanzella2023extremely} observed at $z=6.639$ a potential Pop III system with stellar mass $M_\star <10^4 \msun$, \cite{maiolino2024jades} found at $z= 10.6$ a potential Pop III complex of size $M_\star = 2-2.5 \times 10^5 \msun$ at a distance of around 2 $\kpc$ from its host galaxy GN-z11 \citep{jiang2021evidence, tacchella2023jades} with stellar mass $M_\star \approx 8 \times 10^8 \msun$, \cite{wang2024strong} identified at $z \approx 8.2$ a potential Pop III cluster of mass $M_\star = (7.8 \pm 1.4) \times 10^5 \msun$ at a distance of around 1 $\kpc$ of the main galaxy, and \cite{GLIMPSE2025} found a potential Pop III galaxy candidate at $z\approx 6.5$ with a stellar mass of $\approx 10^5\msun$.

These systems are thought to contain not only Pop III stars but also enriched stars. These observations support the prediction that Pop III star formation at lower redshifts occurs in smaller haloes or satellite galaxies with a spatial offset from the central galaxy. A more detailed comparison of stellar masses would require an explicit model for the evolution of Pop III stars. Such a model would enhance our understanding of the transition from Pop III to Pop II stars and help estimate the timeframe in which Pop III star formation can dominate within individual haloes.

\subsection{Caveats}
\label{subsec:caveats}
The simulations presented in this paper include most of the relevant physical processes for Pop III star formation, including molecular hydrogen chemistry, radiative transfer including Lyman-Werner radiation and a sufficiently high resolution to resolve the smallest star-forming haloes of $\sim 10^6 \msun$.
The latter two points are crucial because our simulation self-consistently models the escape fraction of LW photons eliminating the need to assume a simplified, uniform LW background.
However, some physical processes relevant for Pop III star formation are either absent or not well-constrained in our simulations.

For instance, we do not account for the relative streaming velocity between baryons and dark matter in the early universe. This streaming velocity may reduce the overdensity in small haloes and decays as $\propto (z+1)$. It primarily affects the smallest haloes, thus shifting the critical halo mass for Pop III star formation to higher values. Generally, this effect is relevant for halo masses below $M_{\rm halo} \approx 10^7 \msun$ \citep{schauer2021influence}, which dominate our late-time Pop III star formation.

Additionally, our chemical network does not include deuterium, particularly molecular HD. 
HD formation becomes efficient below $200$\,K due to its lower binding energy compared to molecular hydrogen (chemical fractionation).
These molecules have a non-zero dipole moment and allow cooling below $\approx 100$\,K down to the redshift-dependent CMB temperature floor \citep{nagakura2005formation}. 
The low temperatures required for efficient HD formation can be achieved in various ways—for example, in haloes exposed to external Lyman-Werner radiation \citep{Nishijima2024}, in more massive haloes, or simply through a slower gravitational collapse \citep{greif2011simulations}.
The reduced temperature would lower the Jeans length, potentially leading to earlier star formation in our model.
As we have shown in \cref{fig:birth_subhalo_distribution_3}, Pop III stars form at $z<8$ in subhaloes with masses below $10^8 \msun$, where we expect the impact of HD cooling to be minimal. Moreover, HD formation itself requires cold temperatures and high densities—conditions that, in our model, already lead to star formation.

In our simulations, we use an initial free electron fraction of $10^{-6}$, which is lower than the residual one of $\approx 10^{-4}$ expected after recombination \citep{Galli1998}.
This reduced number of free electrons results in a less efficient formation of molecular hydrogen in minihaloes, potentially shifting the formation of Population III stars to larger halo masses \citep{Latif2025}.
This is especially important at high redshift when there are no external sources present to partially ionize the gas.
This could also explain the shift to lower birth halo masses for Pop III stars at lower redshift, as shown in \cref{fig:birth_subhalo_distribution_3}.

As discussed in more detail in \cref{app:molecularHydrogen}, our chemical network assumes that all \(\mathrm{H}^-\) ions are converted into molecular hydrogen, thereby neglecting destruction by photodetachment and mutual neutralization. As shown in \cref{fig:rates_z16}, this is generally a good approximation, but may break down in dense gas near newly formed Pop~III stars. Nevertheless, since a Pop~III star has already formed in these cases, a more comprehensive chemical network would likely result only in a slight reduction in the overall Pop~III formation rate.

Lyman-Werner radiation can travel through the neutral IGM due to its low opacity below 13.6eV \citep{haiman2000radiative}. As a consequence, shortly after the formation of the first Pop III stars, an extragalactic LW background forms \citep{incatasciato2023modelling}, which can dissociate molecular hydrogen in minihaloes, preventing them from cooling \citep{haiman1997destruction}. Atomic cooling haloes with $M_{\rm halo} >10^7 \msun$ are less sensitive to the LW background, as they have additional cooling channels involving atomic hydrogen, which allows them to reach higher initial densities. Only a powerful local source would be able to prevent them entirely from molecular hydrogen cooling \citep[e.g.][]{regan2017rapid}. 
Our simulations include local LW sources self-consistently, and most of the Pop III star formation happens in haloes with $M_{\rm halo}>10^7 \msun$. 
Nevertheless, we stress that we only inject radiation in the UV bins as a boundary condition and no LW radiation.
To evaluate the strength of local LW sources, we calculate in \cref{app:LWBackground} the mean and median LW intensity in the zoom-in regions as a function of redshift.
In zoom-in regions with more massive halos, the local radiation field exceeds the expected cosmic background, suggesting that adding the latter one would, in these simulations, probably not stop Pop III star formation.
Nevertheless, including an LW background could slightly reduce the Pop III star formation rate in simulations with only a small target halo, though it might be shifted to a lower redshift as haloes surpass the atomic cooling limit.

Our simulations rely on numerical diffusion and assume perfect mixing of metals within resolution elements.
\cite{sarmento2016following} demonstrated that incorporating a subgrid model for metal mixing based on local turbulence can enhance the Pop III star formation rate by up to a factor of 4. This is achieved by allowing gas cells to contain a fraction of pristine gas, even if they are already polluted on average. In a more recent study,
\cite{sarmento2022effects} reported an even greater impact from the subgrid mixing model, up to a factor of 10. They attributed this to their lower resolution compared to their earlier work. However, their use of an AMR grid code makes it nearly impossible to directly compare our resolutions. Nevertheless, at least the dark matter resolution in our lowest resolution simulations is comparable to that in \cite{sarmento2016following}. 
Further investigation into the influence of a subgrid mixing model on our setup will be valuable for quantitative comparisons. 

The main limitation of our model is the absence of an explicit Pop III stellar evolution model that accounts for their top-heavy IMF and stronger stellar feedback. \cite{klessen2023first} showed using the Geneva code \citep{eggenberger2008geneva} that Pop III stars can generate about 100 times more hydrogen-ionizing photons per baryon, with even larger ratios for helium-ionizing photons.
This increased photon production at early times has the potential to escape the host halo, ionizing the neighbouring IGM and thereby inhibiting star formation in that region at later times.

\section{Summary}
\label{sec:summary}
In this paper, we analyzed the metallicity distribution of newborn stars using zoom-in simulations from the \thzoom project, which includes 14 target haloes spanning a mass range from $1.5 \times 10^{8}\msun$ to $9 \times 10^{12} \msun$ at redshift $z=3$.
The high mass resolution allows us to resolve the smallest star-forming haloes with $M_{\rm halo} \approx 10^6 \msun$ by 1500 dark matter particles in our highest resolution simulations. The simulations incorporate essential processes for understanding star formation, such as a multiphase ISM model, on-the-fly radiative transfer, including Lyman-Werner radiation, a self-consistent large-scale radiation field from the parent cosmological box as a boundary condition, dust physics, and a non-equilibrium primordial chemistry network that includes molecular hydrogen.

Generally, younger stars and those in more massive haloes tend to be more metal-rich on average. This trend arises from the more efficient formation of stars at earlier cosmic times, leading to greater metal enrichment in the ISM.
We observe a transition from Pop III to Pop II stars in the star formation history between redshift $z=18$ and $z=12$, which varies as a function of halo mass; the transition occurs later in less massive haloes.
Only haloes with $M_{\rm halo} > 8 \times 10^{10}\msun$ at redshift $z=3$ are capable of producing Pop I stars at that time.

In our simulations, we observe the formation of Pop III stars until $z \approx 5$, coinciding with the end of reionization. In \cref{sec:PopIIIStarFormation}, we focused on the evolution of pristine gas and the formation of Pop III stars.
We find that the IGM surrounding the target haloes contains most of the pristine gas. High-density pristine gas can be found in haloes ranging from $10^6\msun$ to $10^9\msun$ surrounding the target haloes. 
During reionization, these haloes lose their dense gas due to photoevaporation and eventually merge with the target halo. Consequently, most Pop III star particles can be located in small satellite galaxies within the target haloes at lower redshifts.

In an upcoming paper, we will simulate some haloes using an explicit Pop III star formation model that accounts for their top-heavy IMF, stronger feedback per baryon, and higher metal yields. 
This will allow us to study the transition from Pop III to Pop II star formation in the minihaloes we found in our simulations. Understanding this transition is crucial for estimating the likelihood of discovering Pop III-dominated systems at lower redshifts. Additionally, we intend to calculate the global star formation rate densities (SFRD) for the different stellar populations. 
Our simulations' extended zoom-in regions, particularly at higher redshifts, enable a straightforward estimation in various environments. We will compare our results with those from the full \thesanone box to determine the equivalent SFRD for the entire \thesanone simulation.
We will also further study the properties of the more massive atomic cooling haloes shown in \cref{fig:birth_subhalo_distribution_3}, that form Pop III stars and may serve as potential sites for massive black hole formation. 
A particular area of interest will be whether delayed star formation is driven by external radiation or dynamic heating \citep[e.g.][]{Yoshida2003, Wise2019, Mayer2024}.

\section*{Acknowledgements}
The authors thank Yifei Jin, Xin Wang, Ralf Klessen, Sunmyon Chon, James Trussler and Ruediger Pakmor for useful comments and discussions.
The authors gratefully acknowledge the Gauss Centre for Supercomputing e.V. (\url{www.gauss-centre.eu}) for funding this project by providing computing time on the GCS Supercomputer SuperMUC-NG at Leibniz Supercomputing Centre (\url{www.lrz.de}), under project pn29we. 
Support for OZ was provided by Harvard University through
the Institute for Theory and Computation Fellowship.
RK acknowledges support of the Natural Sciences and Engineering Research Council of Canada (NSERC) through a Discovery Grant and a Discovery Launch Supplement (funding reference numbers RGPIN-2024-06222 and DGECR-2024-00144) and York University's Global Research Excellence Initiative. 
EG is grateful to the Canon Foundation Europe and the Osaka University for their support through the Canon Fellowship. LH acknowledges support by the Simons Collaboration on ``Learning the Universe''.

\section*{Data Availability}
All simulation data, including snapshots, group, and subhalo catalogues and merger trees will be made publicly available in the near future. Data will be distributed via \url{www.thesan-project.com}. Before the public data release, data underlying this paper will be shared on reasonable request to the corresponding author.



\bibliographystyle{mnras}
\bibliography{main} 




\appendix

\section{The density and temperature distribution of star-forming gas}
\label{app:densityDistribution}
\begin{figure}
    \centering
    \includegraphics[width=1\linewidth]{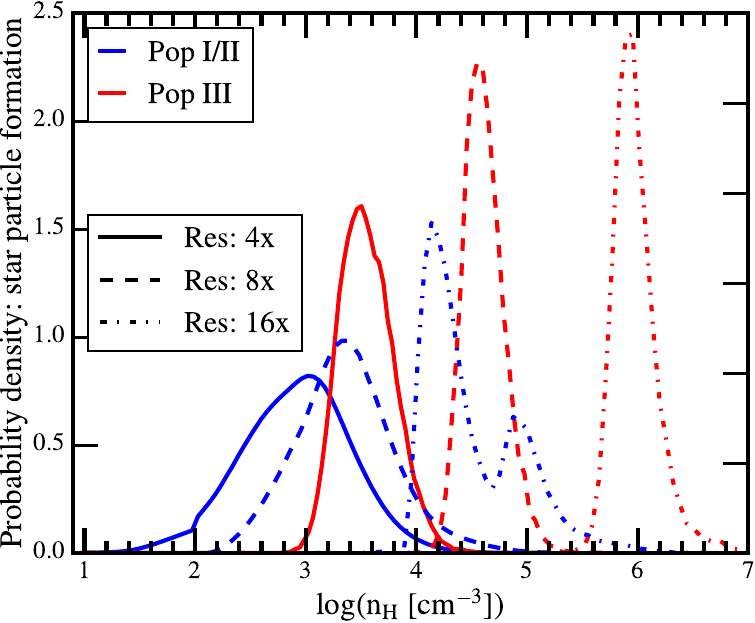}
    \caption{We present the mass-weighted, normalized probability density of the gas density within Voronoi cells immediately prior to the formation of a star particle. Results are shown for three different resolution levels, with higher-resolution cases (8× and 16×) limited to zoom-in simulations of 9 and 5 regions, respectively. The distributions are further separated by stellar metallicity, distinguishing Pop I/II stars (red) from Pop III stars (blue). 
    The distribution shows only a weak dependence on redshift but systematically shifts toward higher densities with increasing resolution. Pop III stars tend to form in denser gas and exhibit a narrower distribution, reflecting the typically higher temperatures and narrower phase structure of their birth environment. }
    \label{fig:stellar_birth_density}
\end{figure}

\begin{figure}
    \centering
    \includegraphics[width=1\linewidth]{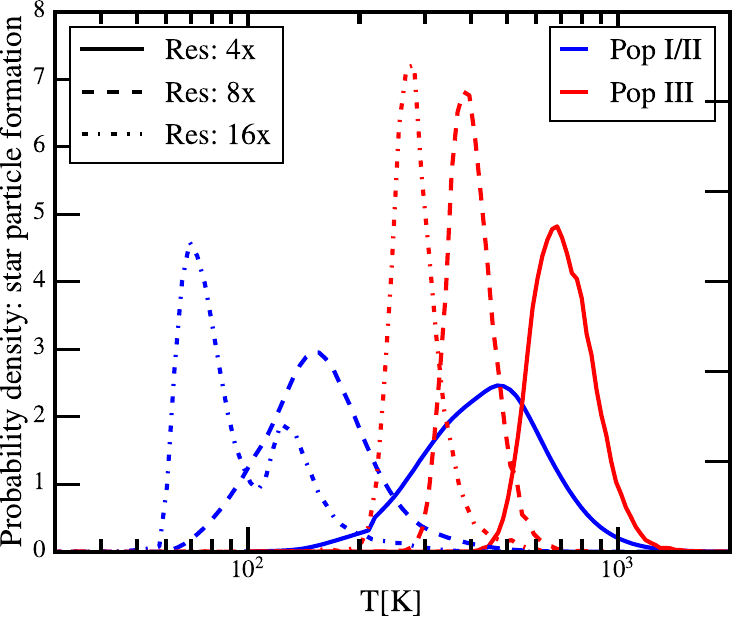}
    \caption{Same as \cref{fig:stellar_birth_density}, but showing the gas temperature at the time of star formation instead of density. Pop III stars form from warmer gas compared to Pop I/II stars. As the resolution increases, the temperature at star formation decreases, consistent with higher gas densities and more efficient cooling prior to star formation.
}
    \label{fig:stellar_birth_temp}
\end{figure}
As discussed in \cref{subsec:methods}, gas cells are eligible to form stars in \thzoom when they become Jeans unstable—that is, when their effective cell size falls below the local Jeans length. Additionally, we impose a density threshold of $n_\mathrm{H} = 10~\mathrm{cm}^{-3}$, to prevent star formation in cold, low-density cells. Within the ISM, we expect the Jeans instability criterion to be the more restrictive of the two. Therefore, in \cref{fig:stellar_birth_density}, we present the density distribution of Voronoi cells immediately prior to the formation of star particles, considering all high-resolution stars in the full zoom-in regions.
All Pop III stars form in gas that is significantly denser than the imposed threshold. As the resolution increases, the typical star formation density also shifts to higher values, which directly reflects the ability to resolve smaller Jeans lengths. Pop III stars typically form in denser gas than Pop I/II stars, as the absence of efficient metal-line cooling leads to higher temperatures in their star-forming regions.
At the highest resolution, the gas density distribution for Pop I/II star-forming cells exhibits a double-peaked structure, with the higher-density peak associated with low-metallicity gas.
In post-processing, we also calculate the temperature distribution of star-forming cells, assuming the birth mass of each star particle corresponds to the Jeans mass of the original gas cell. This assumption is justified by our adoption of a high star formation efficiency. As shown in \cref{fig:stellar_birth_temp}, Pop III stars indeed form in warmer gas, with a lower temperature limit of approximately 200 K—consistent with the minimum temperature achievable through molecular hydrogen cooling. At higher resolutions, we are able to resolve smaller collapsing structures, which in turn allows us to capture lower temperatures during star formation.

\section{The local Lyman-Werner background}
\label{app:LWBackground}
\begin{figure*}
    \centering
    \includegraphics[width=1\linewidth]{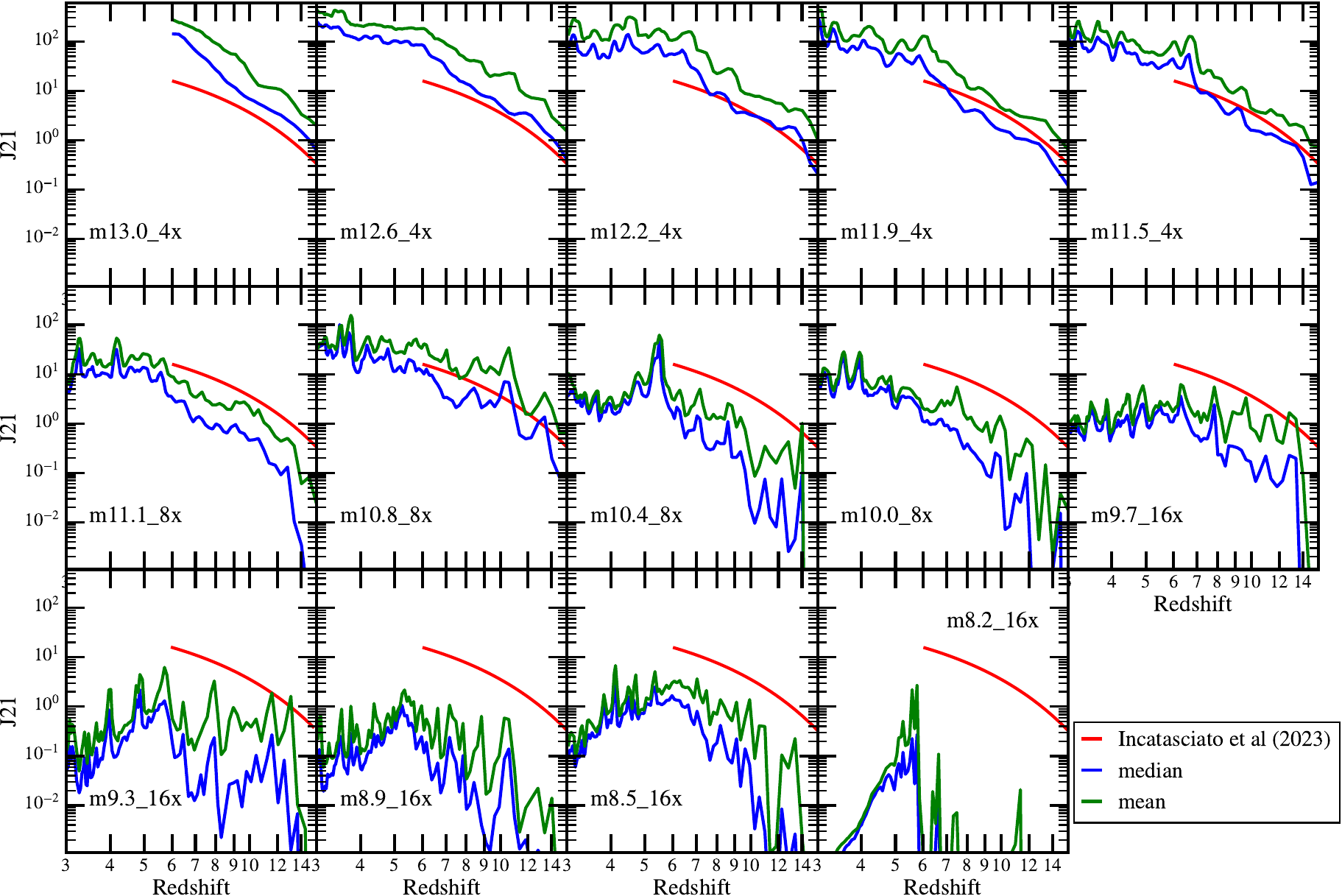}
    \caption{The temporal evolution of the Lyman-Werner radiation intensity in units of $J_{21} = 10^{-21}\mathrm{erg \, s^{-1}\, Hz^{-1}\, sr^{-1}\, cm^{-2}}$ in the high-resolution IGM for all 14 zoom-in regions. 
    We present the median value (blue line), the volume-weighted average (green line), and the fitting function from \protect\cite{incatasciato2023modelling}, which post-processed the FiBY simulation suite and is valid for $z >6$.
    In all simulations, local sources gradually build up an LW background, with this process beginning earlier in zoom-in regions that host larger target haloes.
}
    \label{fig:LW_radiation_all}
\end{figure*}

As discussed in \cref{sec:discussion}, we do not include an external Lyman-Werner (LW) radiation background, considering only local sources. To assess whether these local sources are sufficient to generate an LW background, we calculate the time evolution of the median and volume-weighted mean LW radiation intensity in each zoom-in region, focusing on the high-resolution intergalactic medium (IGM). Here, we define the IGM as all gas cells not associated with any halo.
Consistent with previous studies, we express the intensity in units of \( J_{21} = 10^{-21}\mathrm{erg \, s^{-1} \, Hz^{-1} \, sr^{-1} \, cm^{-2}} \), which is typically calculated at the Lyman limit. For this calculation, we assume a flat spectrum across the LW band and compare our results, shown in \cref{fig:LW_radiation_all}, with the fit from \cite{incatasciato2023modelling}, which is valid for $6 \leq z \leq 23 $. 
As expected, the LW background is stronger in zoom-in regions with more massive target halos. In our representative examples, m12.6\_4x and m10.8\_8x, the mean LW radiation intensity is comparable to or exceeds the predicted cosmological LW background. In contrast, simulations with the smallest target halos show suppressed star formation after reionization, leading to a correspondingly weaker LW background due to the scarcity of local sources.
We note that this calculation was performed using the full zoom-in region, so we expect a higher LW background on average close to the local sources.

\section{The formation of molecular hydrogen in \thzoom}
\label{app:molecularHydrogen}
\begin{figure}
    \centering
    \includegraphics[width=1\linewidth]{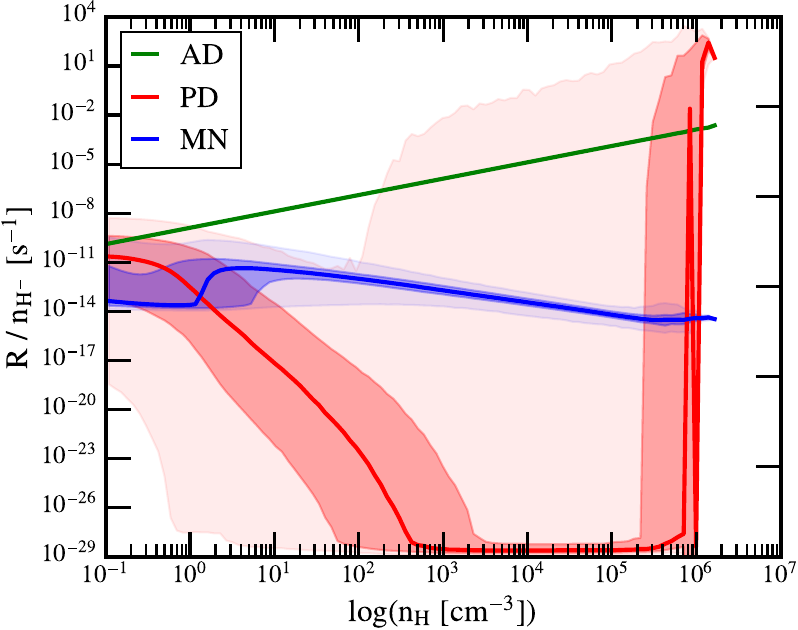}
    \caption{The destruction rates of \(\mathrm{H}^-\) by associative detachment (AD), photodetachment (PD), and mutual neutralization (MN) as a function of density in primordial gas. 
We use simulations with zoom factor 16 and consider gas with metallicity below \(10^{-6} Z_\odot\). 
The solid lines represent the median values, while the shaded regions indicate the 16th--84th and 2nd--98th percentile intervals, respectively. 
Molecular hydrogen formation through AD typically dominates; however, photodetachment can become significant in regions near newly formed stars.
}
    \label{fig:rates_z16}
\end{figure}

\begin{table*}
\centering
\begin{tabular}{ccc}
\hline
\textbf{Coefficient} & \textbf{Value} & \textbf{Source} \\
\hline
$\alpha_{\mathrm{H}_2}^{D}$ & \[
\frac{9.0 \times 10^{-17} \, T_2^{0.5}}{1 + 0.4 T_2^{0.5} + 0.2 T_2 + 0.08 T_2^2} \, \text{cm}^3 \, \text{s}^{-1}
\]& \cite{Nickerson2018}\\
 $\alpha_{\mathrm{H}_2}^{\mathrm{GP}}$ &\[
1.83 \times 10^{-18} \, T_K^{0.88} \, \text{cm}^3\, \text{s}^{-1}
\]& \cite{Hutchins1976, McKee2010}\\
$\alpha_{\mathrm{H}_2}^{\mathrm{3B}}$ & \[
6 \times 10^{-32} T_K^{-0.25} + 2 \times 10^{-31} T_K^{-0.5} \, \text{cm}^6 \, \text{s}^{-1}
\] &\cite{Forrey2013,Nickerson2018}\\

$\sigma_{\mathrm{H}_2 \mathrm{H\,I}}$ & \[
7.073 \times 10^{-19} \, T_K^{2.012} \cdot 
\frac{e^{-5.179 \times 10^4 / T_K}}{\left(1 + 2.130 \times 10^{-5} T_K\right)^{3.512}} 
\, \text{cm}^3\, \text{s}^{-1}
\]& \cite{Dove1986,Nickerson2018}\\
$\sigma_{\mathrm{H}_2 \mathrm{H}_2}$ &\[
5.996 \times 10^{-30} \, T_K^{4.1881} \cdot 
\frac{e^{-5.466 \times 10^4 / T_K}}{\left(1 + 6.761 \times 10^{-6} T_K\right)^{5.6881}} 
\, \text{cm}^3\, \text{s}^{-1}
\]&\cite{Martin1998,Nickerson2018}\\
\hline
\end{tabular}
\caption{The coefficients we use in the chemical network for molecular hydrogen chemistry. The implementation is further discussed in \protect\cite{Kannan2020b}, which itself is based on \protect\cite{Nickerson2018}. $T_K$ is here the temperature in Kelvin and $T_2 = T_K / 100$}
\label{tab:rate_coefficients}
\end{table*}

\thzoom follows the non-equilibrium abundances of \hi, \hii, \HeI, \HeII, \HeIII and \HM.
The reactions governing the ionization and recombination of atomic hydrogen and helium are described in detail in the Appendix B of \citet{Kannan2019}, to which we refer for further information. 
The evolution of the molecular hydrogen abundance is given by:
\begin{align}
\frac{dn_{\mathrm{H}_2}}{dt} =& \alpha_{\mathrm{H}_2}^{D} \left( \frac{D}{D_{\mathrm{MW}}} \right) n_{\mathrm{H}} n_{\mathrm{H\,I}} 
+ \alpha_{\mathrm{H}_2}^{\mathrm{GP}} n_{\mathrm{H\,I}} n_e \\&
+ \alpha_{\mathrm{H}_2}^{\mathrm{3B}} n_{\mathrm{H\,I}}^2 \left(n_{\mathrm{H\,I}} + \frac{n_{\mathrm{H}_2}}{8} \right)
- \sigma_{\mathrm{H}_2 \mathrm{H\,I}} n_{\mathrm{H}_2} n_{\mathrm{H\,I}} \\&
- \sigma_{\mathrm{H}_2 \mathrm{H}_2} n_{\mathrm{H}_2}^2 
- n_{\mathrm{H}_2} \left( S_{\mathrm{H}_2} \Gamma_{\mathrm{H}_2}^{\mathrm{LW}} + \Gamma_{\mathrm{H}_2}^{+} \right),
\end{align}
where $\alpha_{\mathrm{H}_2}^{D}$ is the rate of H$_2$ formation on dust grains, with \( D \) and \( D_{\mathrm{MW}} = 0.01 \) denoting the self-consistently evolved local and Milky Way dust-to-gas ratios, respectively.
The term \( \alpha_{\mathrm{H}_2}^{\mathrm{GP}} \) accounts for low density gas-phase formation, while \( \alpha_{\mathrm{H}_2}^{\mathrm{3B}} \) describes the formation through three-body processes, which become efficient at high densities.
The destruction of molecular hydrogen occurs via collisions with atomic and molecular hydrogen, represented by the rates \( \sigma_{\mathrm{H}_2 \mathrm{H\,I}} \) and \( \sigma_{\mathrm{H}_2 \mathrm{H}_2} \), respectively. The final term describes photodissociation and photoionization. \( \Gamma_{\mathrm{H}_2}^{\mathrm{LW}} \) is the local Lyman--Werner band flux responsible for photodissociation, modulated by the self-shielding factor \( S_{\mathrm{H}_2} \). The self-shielding approximation we adopt follows equation (4) of \citet{Kannan2020b}, based on \citet{Draine1996}, as our radiative transfer scheme does not capture self-shielding by line overlap effects.
The final term $ \Gamma_{\mathrm{H}_2}^{+} $ accounts for photoionization by photons with energies above 15.2eV.
The coefficients used in our chemical model are summarized in \cref{tab:rate_coefficients}.
In dust-free environments, molecular hydrogen formation proceeds exclusively via gas-phase reactions, with the three-body channel becoming efficient only at high densities $n_{\mathrm{H}} > 10^9 \, \text{cm}^{-3}$ \citep{klessen2023first}.
In low-density gas molecular hydrogen can form using $\mathrm{H}^-$ or $\mathrm{H}^+$, with the first one being the dominant one except for rare cases when there are more free protons than electrons or when $\mathrm{H}^-$ photodetachment is much more effective than $\mathrm{H}^+$ photodissociation:
\begin{align}
\mathrm{H} + e^- &\rightarrow \mathrm{H}^- + \gamma  \quad \text{(radiative attachment)}  \\
\mathrm{H}^- + \mathrm{H} &\rightarrow \mathrm{H}_2 + e^-  \quad \text{(associative detachment)} 
\end{align}
Radiative attachment exhibits a significantly lower reaction rate compared to associative detachment.
Therefore, we adopt the reaction rate of radiative attachment as the effective formation rate of molecular hydrogen \citep[similar to ][]{McKee2010}. Rather than explicitly tracking the abundance of $\mathrm{H}^-$, we assume it is in equilibrium between its formation via radiative attachment and its destruction via associative detachment.
This implies that all $\mathrm{H}^-$ is efficiently converted into molecular hydrogen, and that alternative destruction pathways are subdominant. 
This assumption is valid if associative detachment dominates over other destruction mechanisms, which generally include:
\begin{align}
\text{H}^- + \text{H}^+ &\rightarrow \text{H} + \text{H} \quad \text{(Mutual Neutralization)} \\
\text{H}^- + \gamma &\rightarrow \text{H} + e^- \quad \text{(Photodetachment)}.
\end{align}
To assess the relative importance of these three destruction channels, we adopt the following reaction rates:
\begin{align}
R_{\mathrm{AD}} &
\approx \left(1.3 \times 10^{-9}\right) \cdot n_{\mathrm{H}} \cdot n_{\mathrm{H}^-} \quad \text{cm}^{-3}\,\text{s}^{-1} \\
R_{\mathrm{MN}} &
= \left(7 \times 10^{-7} \cdot T^{-0.5} \right) \cdot n_{\mathrm{H}^+} \cdot n_{\mathrm{H}^-} \quad \text{cm}^{-3}\,\text{s}^{-1} \\
R_{\mathrm{PD}} &
= \left(8.8 \times 10^{-10} \cdot J_{21} \right) \cdot n_{\mathrm{H}^-} \quad \text{s}^{-1},
\end{align}
using rate coefficients from \cite{McKee2010}, \cite{Dalgarno1987,Glover2006} and \cite{Omukai2001}, respectively. 
We note that the photodetachment rate is particularly sensitive to the assumed shape of the radiation spectrum \citep{Agarwal2015}.
Assuming charge neutrality, we approximate $n_{\mathrm{H}^+} \approx n_e$.
We compute all \(\mathrm{H}^-\) destruction rates for high-resolution gas with metallicity \( Z < 10^{-6} Z_\odot \), and present the median values as a function of hydrogen number density in \cref{fig:rates_z16}, using our highest-resolution simulations. 
The spread in each rate is shown via the 16th--84th and 2nd--98th percentile intervals.
The rate of associative detachment scales linearly with density and thus exhibits no intrinsic scatter. In contrast, the photodetachment rate depends on the local radiation field. Its median value decreases with increasing density due to enhanced shielding, but rises sharply at \( n_{\mathrm{H}} > 10^5 \, \mathrm{cm}^{-3} \), coinciding with the onset of Pop III star formation in nearby cells (see also \cref{fig:stellar_birth_density}). Photodetachment can dominate \(\mathrm{H}^-\) destruction in low-density gas and even in denser regions (\( n_{\mathrm{H}} \gtrsim 200 \, \mathrm{cm}^{-3} \)) where local Lyman--Werner radiation from newly formed Pop III stars becomes significant.
Mutual neutralization depends on the local free electron density and temperature, and is more efficient at lower temperatures. However, it remains subdominant compared to associative detachment due to the low ionization fractions in our simulations, typically less than \( 10^{-3} \) for gas with \( n_{\mathrm{H}} > 1 \, \mathrm{cm}^{-3} \).
These results are consistent with our lower-resolution simulations, although star formation occurs at slightly lower densities, shifting the density region in which photodetachment becomes the dominant destruction mechanism.
Overall, molecular hydrogen formation remains inefficient in primordial gas, with molecular fractions rarely exceeding 1\%. The median value is typically around \( 10^{-3} \) for gas denser than \( 100 \, \mathrm{cm}^{-3} \).

\section{Additional plots per halo}
\label{app:additionalPlots}
\begin{figure*}
    \centering
    \includegraphics[width=1\linewidth]{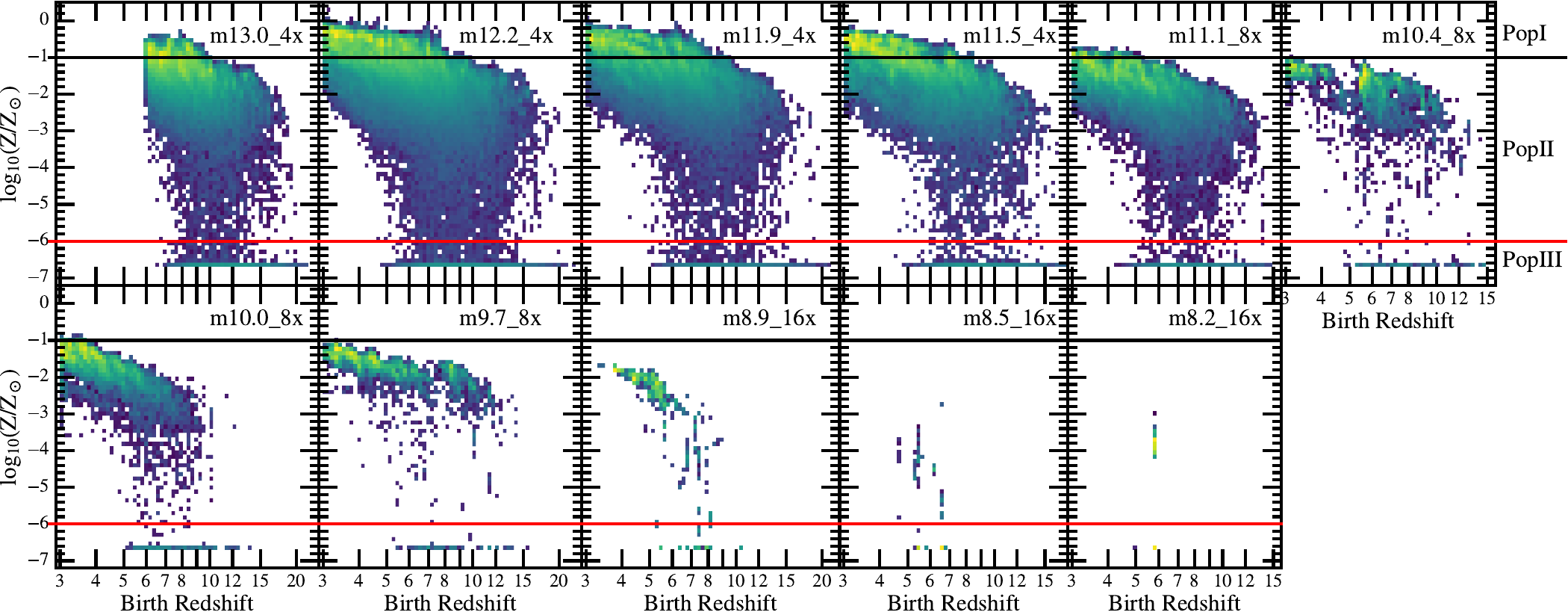}
    \caption{The birth redshift and metallicity of all stars found in the target haloes at redshift $z=3$ ($z=6$ for the most massive one). All haloes contain Pop III stars ($\rm Z < 10^{-6}\,\text{Z}_\odot$) born at the end of the epoch of reionization ($z \approx 5 - 5.5$), which mostly form in satellite galaxies or ex-situ. On average, metal-rich stars can be found in larger haloes and are born at lower redshift. The results for the remaining haloes can be found in \cref{fig:age_metallicity_stars_3}.}
    \label{fig:age_metallicity_stars}
\end{figure*}

\begin{figure*}
    \centering
    \includegraphics[width=1\linewidth]{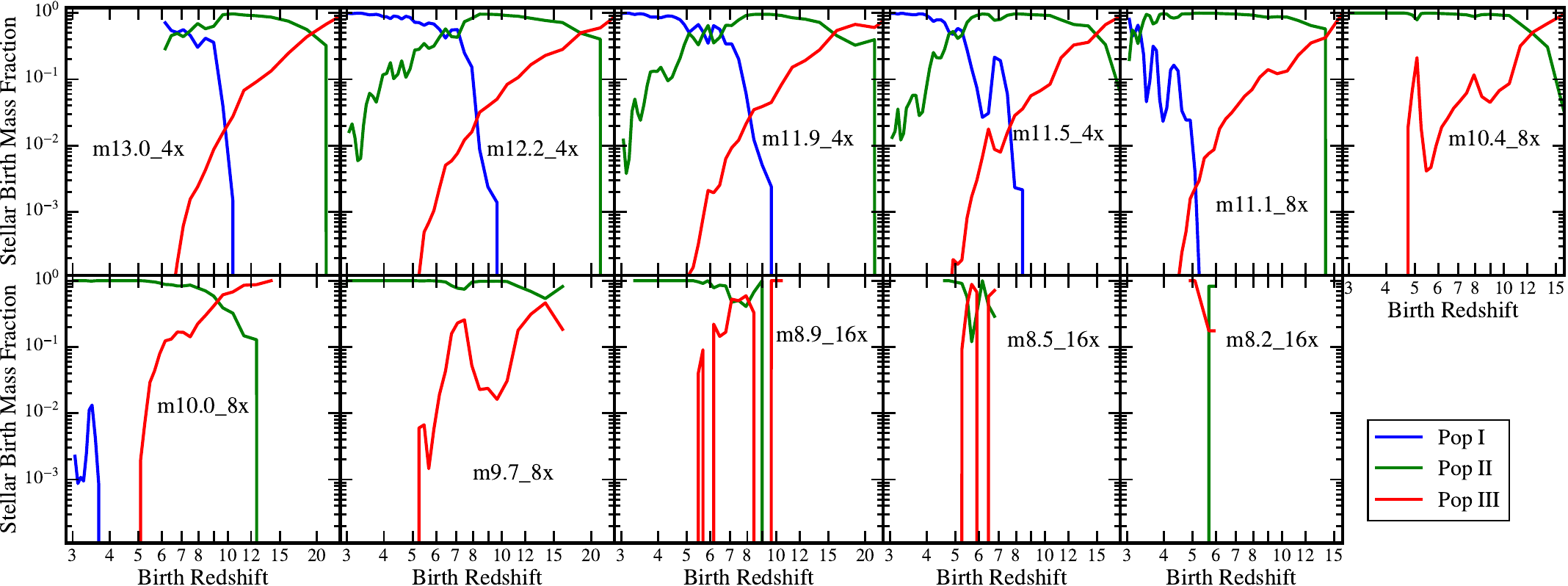}
    \caption{The mass fraction of different stellar populations as a function of the birth redshift. We averaged the results over $100\,\rm Myr$ and used the birth mass of the star particles. We only consider stars in the target halo at $z=3$ ($z=6$ for the most massive halo), but they can be born outside of it.
    This figure is equivalent to \cref{fig:ratio_star_formation_populations_per_halo_3} but contains the additional 11 target haloes. }
    \label{fig:ratio_star_formation_populations_per_halo}
\end{figure*}

\begin{figure*}
    \centering
    \includegraphics[width=1\linewidth]{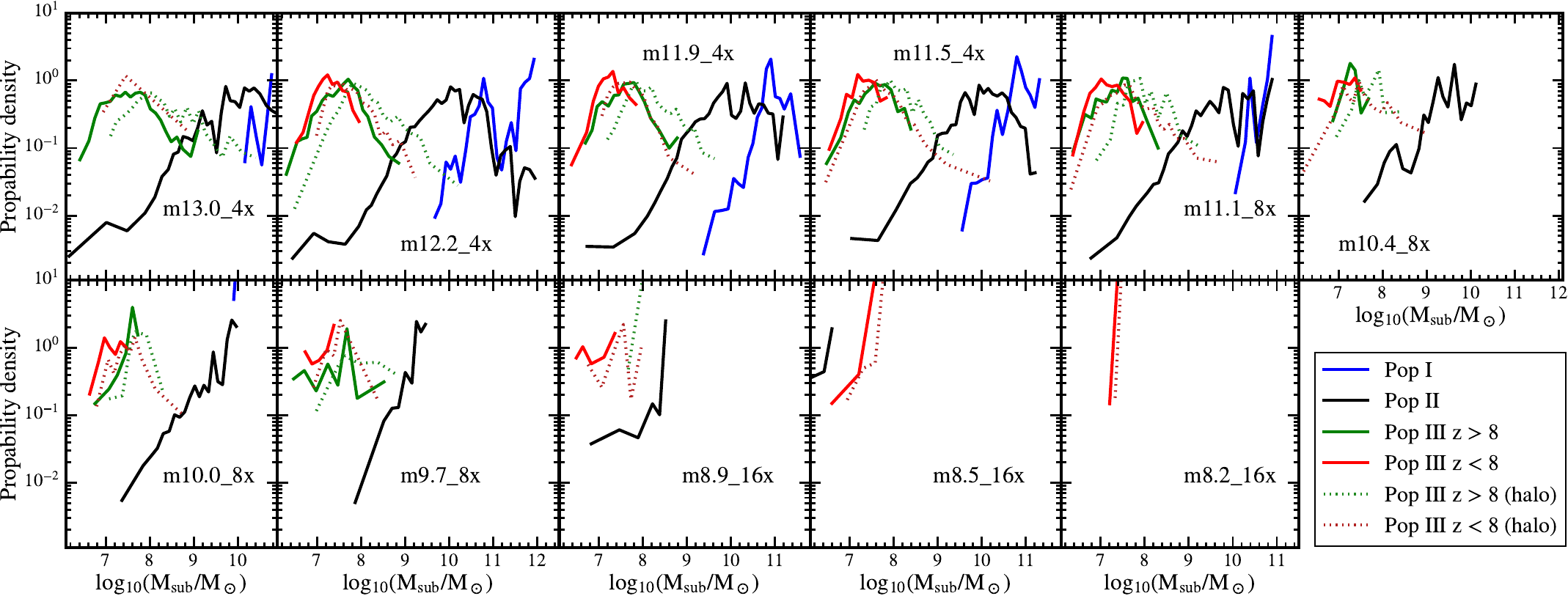}
    \caption{The mass distribution of the subhaloes in which different stellar populations are born. We use all stars that can be found at redshift $z=3$ ($z=6$ the large halo) in our main haloes. 
    We divided Pop III stars into those born before (green line) and after $z=8$ (red line) and show additionally their birth halo mass distribution (dotted lines). 
    Pop I and Pop II stars mostly follow their host halo mass distribution with time, while Pop III stars are almost exclusively formed in subhaloes $\rm <10^8\msun$. 
The data for the remaining three haloes can be found in \cref{fig:birth_subhalo_distribution_3}.}
    \label{fig:birth_subhalo_distribution_full}
\end{figure*}

\begin{figure*}
    \centering
    \includegraphics[width=1\linewidth]{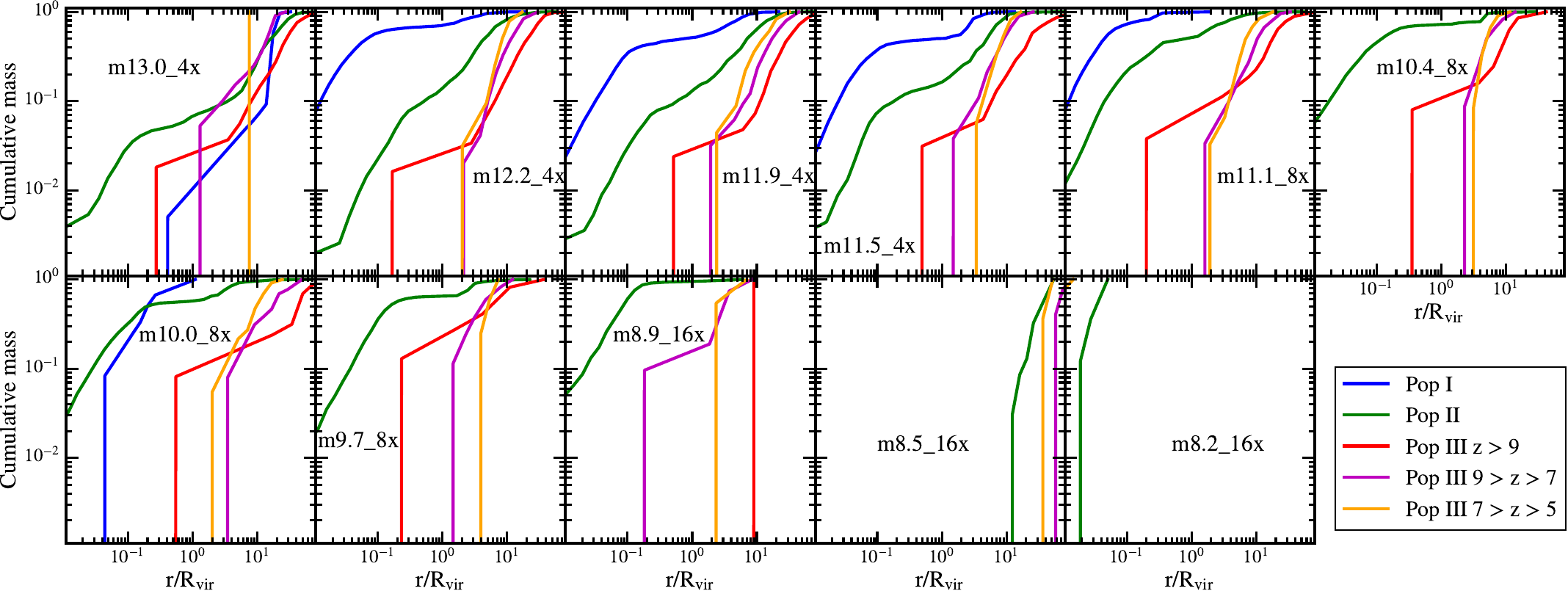}
    \caption{Cumulative mass distribution as a function of distance to the target halo centre at the redshift of birth for different stellar populations and birth redshift ranges. 
    The distances are normalized by the virial radius of the target halo at the redshift of their birth time.
    Except for the largest halo that grows through mergers, Pop I and Pop II stars mostly form within the virial radius, while Pop III stars after $z=9$ form almost exclusively outside of the virial radius. The data for the remaining three target haloes can be found in \cref{fig:radial_mass_distribution_origin_cum_3}.}
    \label{fig:radial_mass_distribution_origin_cum_rel_r_vir}
\end{figure*}

For a clearer representation, we focused on three representative haloes in the main body of this paper. This Appendix displays the main data for the remaining haloes of the \thzoom project, which exhibit the same qualitative behaviour.
We show in \cref{fig:age_metallicity_stars} the stellar age-metallicity relation, in \cref{fig:ratio_star_formation_populations_per_halo} the temporal evolution of the mass fraction of different stellar populations, in \cref{fig:birth_subhalo_distribution_full} the mass distribution of the birth subhaloes for different stellar populations, and in \cref{fig:radial_mass_distribution_origin_cum_rel_r_vir} the cumulative mass distribution of different stellar populations as function of their distance to the centre of the primary halo at the time of their birth.

\bsp	
\label{lastpage}
\end{document}